\def\CLASSINPUTbaselinestretch{1.025}
\documentclass[11pt,journal,compsoc,onecolumn,a4paper]{IEEEtran}
\renewcommand{\IEEEcompsocdiamondline}{}

\usepackage[T1]{fontenc}
\usepackage[utf8]{inputenc}

\usepackage{amsmath}
\usepackage{amssymb}
\usepackage{amsfonts}
\usepackage{dsfont}
\usepackage{xfrac}

\usepackage{graphicx}
\usepackage[table]{xcolor}
\usepackage{booktabs}
\usepackage{tabularx}
\usepackage{multirow}
\usepackage{makecell}
\usepackage{wrapfig}
\usepackage{float}
\usepackage{tcolorbox}

\usepackage{subfig}
\usepackage{caption}

\DeclareCaptionLabelSeparator{towardspipe}
{\hspace{0.45em}|\hspace{0.45em}}

\usepackage[procnumbered,ruled]{algorithm2e}
\usepackage{enumitem}
\usepackage{balance}

\usepackage{textcomp}
\usepackage[normalem]{ulem}

\usepackage{hyperref}
\hypersetup{
    hidelinks
}

\newcommand{\inlinebulletsection}[1]{%
    \par\medskip
    \noindent
    \textbullet\enspace
    \textbf{#1}\enspace
    \ignorespaces
}

\newcolumntype{P}[1]{%
    >{\centering\arraybackslash}p{#1}%
}

\newcolumntype{L}[1]{%
    >{\raggedright\arraybackslash}p{#1}%
}

\makeatletter
\newcommand{\thickhline}{%
    \noalign{\ifnum0=`}\fi
    \hrule height 1pt
    \futurelet\reserved@a\@xhline
}
\makeatother

\usepackage[
    a4paper,
    left=25mm,
    right=25mm,
    top=25mm,
    bottom=25mm
]{geometry}

\usepackage{titlesec}

\titleformat{\section}
    {\normalfont\rmfamily\bfseries
     \fontsize{13bp}{15.5bp}\selectfont}
    {\thesection.}
    {0.5em}
    {}

\titleformat{\subsection}
    {\normalfont\rmfamily\bfseries
     \fontsize{11bp}{13.5bp}\selectfont}
    {\thesubsection.}
    {0.5em}
    {}

\titleformat{\subsubsection}
    {\normalfont\rmfamily\bfseries
     \fontsize{11bp}{13.5bp}\selectfont}
    {\thesubsubsection.}
    {0.5em}
    {}

\titlespacing*{\section}
    {0pt}{1.8ex plus 0.5ex minus 0.2ex}{0.8ex}

\titlespacing*{\subsection}
    {0pt}{1.5ex plus 0.4ex minus 0.2ex}{0.6ex}

\titlespacing*{\subsubsection}
    {0pt}{1.3ex plus 0.4ex minus 0.2ex}{0.5ex}

\usepackage{ragged2e}\newenvironment{reportabstract}{%
    \par
    \vspace{-0.5\baselineskip}
    {%
        \centering
        \normalfont
        \rmfamily
        \bfseries
        \fontsize{16bp}{19bp}\selectfont
        Abstract
        \par
    }%
    \vspace{0.4\baselineskip}
    \begingroup
    \normalfont
    \rmfamily
    \fontsize{10.95bp}{13.55bp}\selectfont\justifying
    \setlength{\parindent}{0pt}%
    \setlength{\parskip}{0pt}%
}{%
    \par
    \endgroup
}

\newcommand{\reportkeywords}[1]{%
    \par
    \vspace{0.8\baselineskip}
    \begingroup
    \normalfont
    \rmfamily
    \fontsize{10.95bp}{13.55bp}\selectfont\justifying
    \setlength{\parindent}{0pt}%
    \noindent
    \textbf{Keywords:}\enspace #1
    \par
    \endgroup
}

\renewcommand{\footnoterule}{%
    \kern -3pt
    \noindent\makebox[\columnwidth][c]{%
        \rule{1\columnwidth}{0.2pt}%
    }%
    \kern 2.6pt
}

\title{%
    \rmfamily\bfseries
    \fontsize{16bp}{19bp}\selectfont
    From Ranked Documents to Reliable Contexts: 
    An Answer-Oriented Context Construct Framework for AI Search
}

\author{
{\large\rmfamily
Yunfei Zhong, Yinqiong Cai, Lixin Su, Haosheng Qian, Lixin Zou,\\[2pt]
Yixing Fan, Sheng Xu, Jiafeng Guo, Daiting Shi, and Jingzhou He
}%
\IEEEcompsocitemizethanks{
\IEEEcompsocthanksitem
Yunfei Zhong, Yixing Fan, and Jiafeng Guo are with the
State Key Laboratory of AI Safety, Institute of Computing Technology,
Chinese Academy of Sciences, Beijing, China, and with the
University of Chinese Academy of Sciences, Beijing, China
(e-mail: \texttt{zhongyunfei25s@ict.ac.cn}).

\IEEEcompsocthanksitem
Yinqiong Cai, Lixin Su, Haosheng Qian, Sheng Xu, Daiting Shi and Jingzhou He are with
Baidu Inc., Beijing, China. Yinqiong Cai is the corresponding author
(e-mail: \texttt{yinqiongcai@gmail.com}).

\IEEEcompsocthanksitem
Lixin Zou is with Wuhan University, Wuhan, China.
}
}

\begin{document}

\IEEEtitleabstractindextext{%
\begin{reportabstract}

Traditional Web search follows a human-facing paradigm in which ranked documents are presented directly to users, who inspect the results and synthesize the needed information on their own. 
With large language models (LLMs), AI Search follows a model-facing paradigm in which retrieved documents are first selected and organized into a bounded context, which is then consumed by a generation model to produce the final answer.
This shift changes the retrieval objective from ranking individual documents according to Search Satisfaction to constructing a reliable context for consistent and robust correct answer generation.

To operationalize this shift, we reformulate retrieval in AI Search as answer-oriented context construction and introduce a three-stage framework:
(1) \emph{Answer Support} determines whether candidate documents contain answer-bearing information that can contribute to answer generation;
(2) \emph{Content Trustworthiness} determines whether this information is sufficiently reliable to support correct answer generation from source, temporal, and factual perspectives;
and (3) \emph{Context Organization} jointly selects, consolidates, and structures the retained information under a finite context budget to consistently and robustly generate correct answers.

Building on this framework, we develop an industrial AI Search workflow spanning both prior and posterior optimization. Prior optimization improves the pre-generation context through the three-stage framework, while posterior optimization uses answer-level feedback to attribute generation outcomes to upstream document and context-organization decisions. We further establish a systematic evaluation protocol that assesses both the retrieval-side context supplied to the generation model and the final answers delivered to users. Experiments show that the proposed upgrades consistently improve performance at both the Retrieval and Answer levels, demonstrating the effectiveness of the proposed framework and its industrial implementation.

\end{reportabstract}

\reportkeywords{%
Information Retrieval, Web Search, AI Search, Retrieval-Augmented Generation, Large Language Models, Trustworthy Retrieval, Context Construction
}
}

\maketitle
\thispagestyle{empty}

\newpage
\tableofcontents
\newpage

\section{Introduction}
\label{sec:introduction}

Traditional Web search is built around a human-facing ranked-list interface, in which retrieval and ranking systems place the documents most likely to meet an information need near the top of the result page \cite{brin1998anatomy,yang2005webir}. Given a query, the system retrieves a set of candidate documents, applies an initial ranking followed by one or more reranking stages, and presents the ranked list of URLs to the user; then users decide which results to click, read the pages, compare sources, and form their own conclusions \cite{wilson1981userstudies,dervin1983sensemaking,rieh2002judgment,rieh2007credibility,schwarz2011augmenting}. 
Commercial search engines model this objective through a \emph{Search Satisfaction} score, which evaluates candidate documents along the axes of Relevance, Authority, Freshness, and Quality: Relevance captures basic satisfaction, while Authority, Freshness, and Quality characterize high-quality satisfaction.
By modeling the interaction between the query and each document, Search Satisfaction fulfills the system's responsibility of helping users find promising results, while reading, comparison, verification, and final synthesis remain under the user's control\cite{rieh2002judgment,rieh2007credibility,schwarz2011augmenting}.

\begin{figure}[H]
    \centering
    \includegraphics[
        width=1\linewidth
    ]{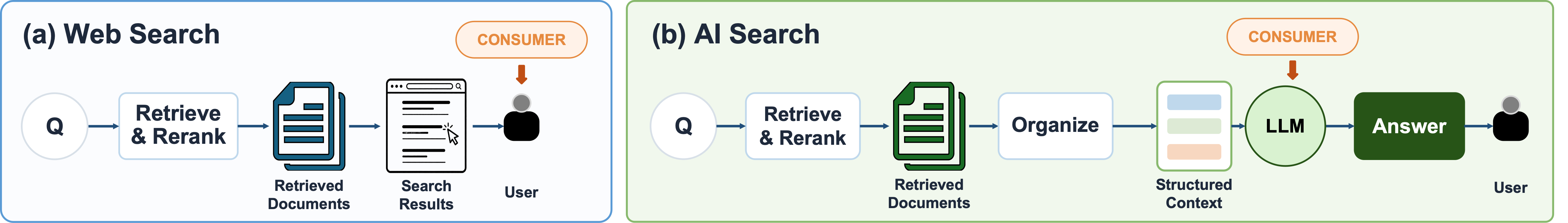}
    \caption{Comparison of human-facing Web Search and model-facing AI Search. Web Search presents a ranked list of documents to users, whereas AI Search organizes retrieved documents into a structured context for LLM-based answer generation.}
    \label{fig:intro-compare}
\end{figure}

AI search changes this information-seeking paradigm. In the retrieval-augmented generation (RAG) pipeline, web pages are no longer presented directly to humans, but instead serve as intermediate inputs to a large language model that generates the final answer \cite{lewis2020rag,gao2024ragsurvey,fan2024ragsurvey, izacard2021leveraging}. As illustrated in Fig.~\ref{fig:intro-compare}, the primary consumer of the retrieved information shifts from the human user to the generation model, and subsequently the user directly consumes the resulting answers\cite{zhang2026beyond}.  
Before the answer is generated, the search system must determine which retrieved documents to provide as evidence and how the evidence is structured as context to the generator.
In other words, the information refinement task that is completed by users in the traditional Web search is now addressed in the AI search engine.

This transition creates a structural mismatch between Search Satisfaction and the requirements of AI Search.
First, Relevance measures how well a document matches the query in terms of topic, intent, and semantics, but such alignment does not necessarily indicate its contribution to answer construction. A page that is relevant to the query may provide little of the information needed to derive the answer; conversely, a less directly relevant document may supply facts or premises that make a substantive contribution to answer construction \cite{mao2016relevance,joren2024sufficient,lee2025setselection,salemi2024searchmachines,yu2024rankrag}.
Second, Authority, Freshness, and Quality provide human-facing cues about whether a result comes from an authoritative source, is sufficiently recent, and is well organized for consumption. These cues help users judge whether the information is worth inspecting and adopting, but they do not fully establish whether the information itself can be relied upon for model consumption. In particular, source authority does not guarantee that the source is appropriate for the specific information being used, publication recency does not ensure temporal validity, and well-presented content may still contain factually unreliable claims. This limitation becomes more consequential in AI Search because users primarily observe the generated answer rather than directly inspecting the underlying evidence. Source-inappropriate, temporally invalid, or factually unverified information entering the model context may therefore be propagated into a misleading answer that is harder for users to detect than an unreliable individual search result \cite{hwang2024rarag,jin2026sourcebench,ouyang2025hoh,zhang2021situatedqa,chen2026expiration,thorne2018fever,min2023factscore}.
Third, traditional Web search ranks retrieved documents according to their Search Satisfaction scores and presents the resulting ranked list to users. This representation is suitable for human consumption because users can decide which results to inspect, compare information across documents, and synthesize the needed information themselves. However, directly treating the ranked list as the context supplied to the generation model is not necessarily optimal. Even when the retrieved documents collectively contain the information required for a correct answer, irrelevant content, cross-document redundancy, and evidence placement can hinder the model from using that information appropriately \cite{yoran2023irrelevant,liu2024lost}. AI Search must therefore determine not only which documents to retain, but also what information from them should enter the model context and how that information should be organized for generation.
To address these gaps, we extend the retrieval objective beyond identifying satisfactory individual documents to constructing a reliable and model-usable context.
Accordingly, we propose a three-stage framework for assessing retrieved documents and organizing the information into a reliable context for consistently and robustly generating accurate answers:

\begin{enumerate}[leftmargin=1.8em]
    \item \emph{Stage 1: Answer Support} addresses the limitation that query--document relevance does not necessarily indicate whether a document contributes information to answer generation. It is a query-conditioned, document-level criterion that goes beyond topical relevance or exact term matching by evaluating whether a candidate document is answer-bearing or provides indirect evidence from which the answer can be derived \cite{joren2024sufficient}. Its goal is to ensure that the retained documents contain the information needed for answer generation.
    
    \item \emph{Stage 2: Content Trustworthiness} addresses the mismatch between human-facing satisfaction signals and the reliability requirements of model consumption. It is a document-level criterion that evaluates whether the answer-supporting information can be relied upon for accurate answer generation along three dimensions.
    \emph{Source Trustworthiness} assesses whether the information originates from a source with the authority, expertise, or first-hand access required for the information it provides, such as an authoritative institution or a recognized expert publisher in the relevant domain. \emph{Temporal Trustworthiness} assesses whether the information remains applicable to the time, version, policy state, or event stage required by the query and has not expired or been superseded. \emph{Information Trustworthiness} assesses whether the factual claims expressed in the document can be verified and corroborated by traceable evidence. Together, these dimensions aim to ensure that the retained information can serve as a reliable basis for generating a correct answer.

    \item \emph{Stage 3: Context Organization} addresses the shift of information selection and organization from user-side interaction to pre-generation context construction. It is a set-wise process that determines what information from the retained documents should be included in the model input and how that information should be structured and ordered under a finite context budget. This process removes irrelevant content within documents, deduplicates overlapping information across documents, and organizes the remaining content into structured context for consistent, robust, and accurate answer generation.
\end{enumerate}

To operationalize the proposed framework, an industrial implementation workflow comprises prior and posterior optimization. Prior optimization operates before answer generation and instantiates the three stages through \emph{LLM-based Answer-Supporting Document Ranking}, \emph{Document Trustworthy Degree Assessment}, and \emph{Set-wise Document Selector and Context Organization}. Posterior optimization operates on the generated answer and addresses the shift from document-level behavioral feedback in Web search to sparse answer-level feedback in AI Search. Because such feedback cannot directly reveal which preceding retrieval or context-construction decisions caused the observed outcome, we implement a \emph{posterior-optimization simulator} that uses model-based proxy judgments to evaluate answers and attribute answer quality back to contributing source documents.

We further establish a systematic evaluation protocol aligned with the pre- and post-generation stages of the optimization workflow. The protocol evaluates two objects: the \emph{retrieval results} supplied to answer generation and the \emph{generated answers} delivered to users. For scalable evaluation, an LLM-based evaluator applies standardized rubrics to both objects, enabling broad coverage and efficient system iteration. Human evaluation complements the automatic assessment at the answer level, where annotators compare generated answers from the perspective of end users and judge which better satisfies the information need. We further define corresponding evaluation metrics for each setting. Online experiments show that the proposed component upgrades consistently improve performance, enhancing both the information supplied to the generation model and the final answers delivered to users.

In summary, this work makes the following contributions:

\begin{itemize}[leftmargin=1.6em]
    \item We characterize the shift from traditional Web search to AI Search, where retrieved documents are no longer consumed directly by users but instead serve as inputs to the generation model. This shift reveals that the human-facing Search Satisfaction objective does not fully capture the requirements of constructing reliable generation context.
    
    \item We propose a three-stage, answer-oriented framework for assessing retrieved information and organizing it into a reliable context for answer generation, comprising \emph{Answer Support}, \emph{Content Trustworthiness}, and \emph{Context Organization}. We further operationalize these objectives in a production workflow through prior optimization before answer generation and posterior optimization based on answer-level outcomes.
    
    \item We establish a systematic evaluation protocol and corresponding metrics for both the retrieval results supplied to the generation model and the generated answers delivered to users. Online experiments on production component upgrades demonstrate consistent improvements at both levels, validating the effectiveness of the proposed framework and optimization workflow.
\end{itemize}

\section{Preliminary and Problem Formulation}
\label{sec:framework}

As discussed in Section~\ref{sec:introduction}, once retrieved information is consumed by the generation model rather than directly by users, the system's responsibility extends beyond ranking individual documents to constructing the context used for answer generation. This section formalizes this shift in two steps. We first formulate \emph{Search Satisfaction}, the query--document modeling objective used in traditional Web search, and clarify its evaluation scope. We then identify the additional requirements introduced by model consumption and formulate a three-stage framework for constructing reliable generation context.

\subsection{Satisfaction-Oriented Web Search}
\label{subsec:satisfaction-ranking} 

In traditional human-facing search, we use a \emph{Search Satisfaction} framework to estimate whether a candidate document could satisfy the user's information need. Let $q$ denote a query, $d$ a candidate document, and $t_0$ the search time. Search Satisfaction is defined as
\begin{equation}
S_{\mathrm{sat}}(q,d;t_0)
=
\Phi_{\theta}\!\left(
\underbrace{
R(q,d)
}_{\text{basic satisfaction}}
,
\underbrace{
A(q, d),
T(q,d;t_0),
Q(q, d)
}_{\text{high-quality satisfaction}}
\right),
\label{eq:search-satisfaction}
\end{equation}
where $S_{\mathrm{sat}}(q,d;t_0)$ denotes the satisfaction score assigned to document $d$ for query $q$ at search time $t_0$.
The four signals---Relevance $R(q,d)$, Authority $A(q,d)$, Freshness $T(q,d;t_0)$, and Quality $Q(q,d)$---reflect different conditions under which a search result can satisfy a user and are defined as follows.

\inlinebulletsection{Relevance.}
$R(q,d)$ measures the human-perceived alignment between the query and the document. The judgment considers how well the document matches the query in terms of subject, information need and explicit constraints, considering both semantic similarity and lexical correspondence. Relevance determines whether the document is a potential candidate to satisfy the query.

\inlinebulletsection{Authority.}
$A(q, d)$ measures the authority level assigned to the source site of the document $d$. Let $s_d$ denote the source category under a predefined taxonomy that distinguishes, for example, government websites, enterprise websites, and verified individual publishers. Authority is defined as
\begin{equation}
A(q,d)
=
g_A(s_d),
\label{eq:authority}
\end{equation} 
where $g_A$ is a rule-based mapping from the predefined site category $s_d$ to an ordinal authority level. Under this formulation, the Authority signal is determined solely by the source site and is independent of the specific query.

\inlinebulletsection{Freshness.}
$T(q,d;t_0)$ measures whether document $d$ satisfies the freshness requirement of query $q$ at search time $t_0$. Let $f_q$ denote the query-side freshness-demand level and $p_d$ the publication time of the document. Freshness is defined as
\begin{equation}
T(q,d;t_0)
=
g_T(f_q,p_d;t_0),
\label{eq:freshness}
\end{equation} 
where $g_T$ compares the age of the document at time $t_0$ with the recency requirement associated with $f_q$. The resulting signal estimates whether the document should be considered outdated for the query based on its publication time.

\inlinebulletsection{Quality.}
$Q(q,d)$ measures the quality of the document's organization and presentation for direct human consumption. It characterizes how effectively the document structures and presents its content so that users can locate and understand the needed information. Unlike Relevance and Freshness, this property is determined primarily by the document itself rather than by its relation to the query. Quality is defined as
\begin{equation}
Q(q,d)
=
g_Q(o_d),
\label{eq:quality}
\end{equation} 
where $o_d$ denotes the document-side organization and presentation features of document $d$, such as structural clarity, readability, and the organization of its content, and $g_Q$ maps these features to the document-level Quality score.

The fusion function $\Phi_{\theta}$ combines the four signals with query-dependent contributions. Their importance varies because different queries impose different conditions for satisfaction. For example, Authority may receive greater weight for queries seeking official information, whereas Freshness may be emphasized for queries concerning rapidly changing events.

\begin{figure}[H]
    \centering
    \includegraphics[
        width=1\linewidth,height=0.3\textheight
    ]{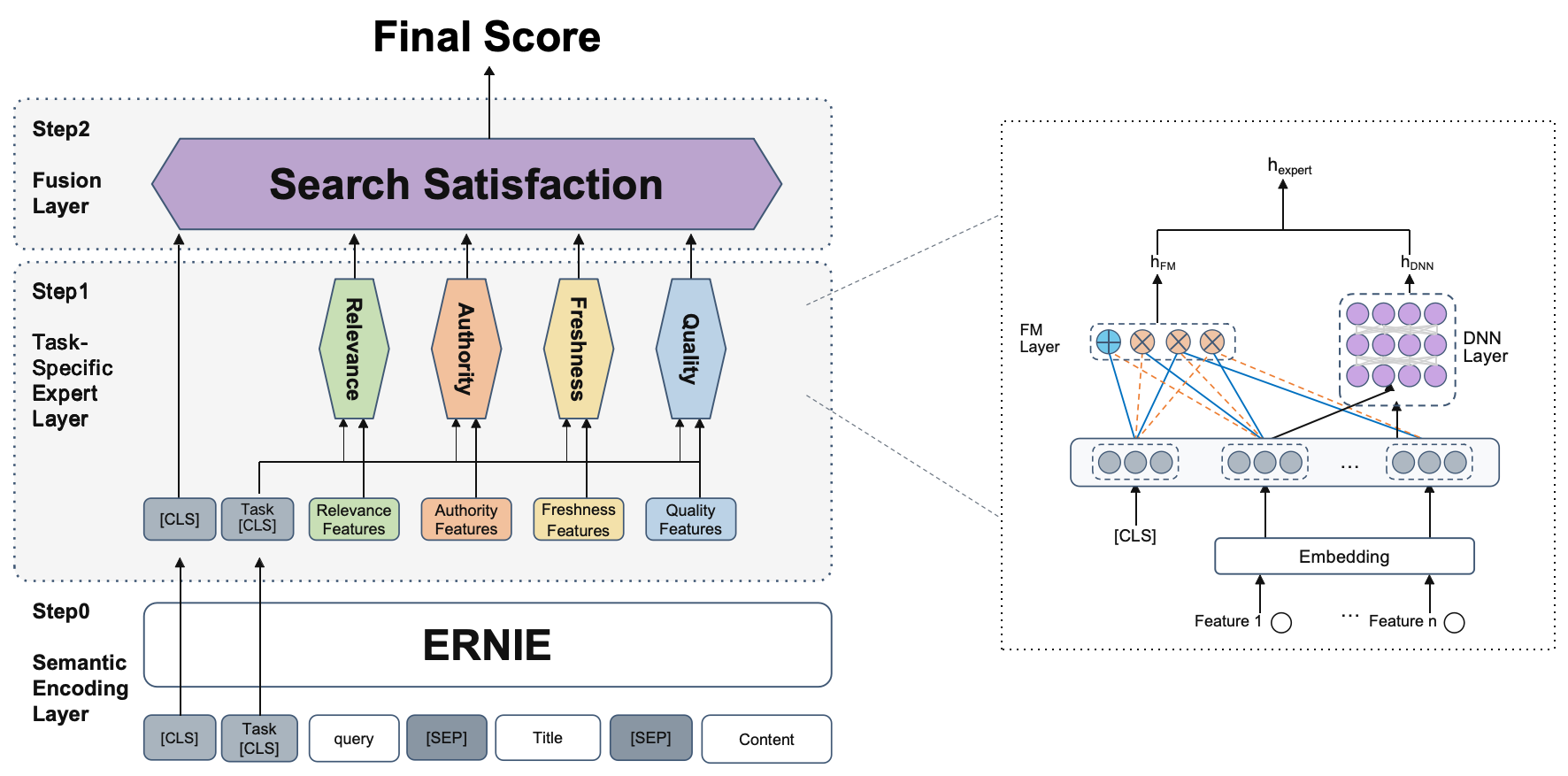}
    \caption{Overall Framework of Search Satisfaction for Web Search.}
    \label{fig:search-satisfaction-model}
\end{figure}

Figure~\ref{fig:search-satisfaction-model} illustrates the implementation architecture of the Search Satisfaction framework.
A semantic encoder jointly represents the query, title, and document content, producing a global representation $\mathbf{h}_{\mathrm{cls}}$ and task-oriented representations $\mathbf{h}_{\mathrm{task}}$ for each business objective:
\begin{equation}
\left(
\mathbf{h}_{\mathrm{cls}},
\left[
\mathbf{h}^{\mathrm{task}}_k
\right]_k
\right)
=
\operatorname{Enc}_{\psi}
\!\left(
q,d_{\mathrm{title}},d_{\mathrm{content}}
\right),
\end{equation} 
where $\mathbf{h}_{\mathrm{cls}}$ denotes the global semantic representation, $\mathbf{h}^{\mathrm{task}}_k$ denotes the task-oriented representation associated with expert $k$, and $\left[\mathbf{h}^{\mathrm{task}}_k\right]_k$ denotes the collection of task-oriented representations over all experts.
Each business-specific expert combines its corresponding task-oriented representation with task-specific features $\mathbf{x}_k$:
\begin{equation}
\mathbf{h}_{k}
=
E_k\!\left(
\mathbf{h}^{\mathrm{task}}_k,
\mathbf{x}_{k}
\right),
\qquad
k\in
\{
\mathrm{rel},
\mathrm{auth},
\mathrm{fresh},
\mathrm{qual}
\},
\end{equation} 
where $\mathbf{h}_k$ is the representation produced by expert $k$, corresponding respectively to Relevance, Authority, Freshness, and Quality.
The resulting expert representations are then fused with the global representation $\mathbf{h}_{\mathrm{cls}}$ by the satisfaction head to produce the final ranking score:
\begin{equation}
S_{\mathrm{sat}}(q,d;t_0)
=
H_{\mathrm{sat}}\!\left(
\mathbf{h}_{\mathrm{cls}},
\left[
\mathbf{h}_{k}
\right]_{k}
\right),
\end{equation} 
where $[\mathbf{h}_{k}]_{k}$ denotes the ordered collection of expert representations corresponding to the four business objectives: $k\in\{\mathrm{rel},\mathrm{auth},\mathrm{fresh},\mathrm{qual}\}$.

In the implementation shown in Fig.~\ref{fig:search-satisfaction-model}, each expert jointly models semantic representations and task-specific feature interactions. 
The expert adopts a DeepFM-style architecture \cite{guo2017deepfm}, in which the FM branch captures explicit low-order feature interactions, while the DNN branch models implicit high-order interactions.
Their outputs are combined to form the business-specific expert representation, which is then fused with the global semantic representation by the satisfaction head to produce the final Search Satisfaction score.
Let $\mathcal{C}_{\mathrm{sat}}(q)$ denote the candidate document set evaluated by the human-facing Web search pipeline for query $q$. The resulting Search Satisfaction score is used to rank these candidate documents:
\begin{equation}
\pi_{\mathrm{human}}(q;t_0)
=
\operatorname*{argsort}_{d\in\mathcal{C}_{\mathrm{sat}}(q)}
S_{\mathrm{sat}}(q,d;t_0)
\qquad
\text{in descending order}.
\end{equation}

In summary, Search Satisfaction models human-facing ranking at the query-document level. It evaluates whether a document could satisfy the user's information need. Its evaluation unit is a single query-document pair, and its output represents the document's suitability as a search result for the query based on the four Search Satisfaction signals: Relevance, Authority, Freshness, and Quality.

\subsection{Answer-Oriented AI Search}

The preceding formulation is designed for human-facing Web search, where documents in $\mathcal{C}_{\mathrm{sat}}(q)$ are evaluated individually and ranked according to their Search Satisfaction scores for direct user inspection.
In AI Search, however, the ranked documents are not directly consumed by users. Instead, candidate documents undergo further assessment before being used for answer generation. Let $\mathcal{C}_{\mathrm{ret}}(q)$ denote the candidate documents retained after this assessment. These documents are then organized into a generation context:
\begin{equation}
\mathbf{X}
=
O\!\left(q,\mathcal{C}_{\mathrm{ret}}(q);B\right),
\qquad
|\mathbf{X}|\le B,
\label{eq:context-organization}
\end{equation} 
where $O$ denotes the context organization process and $B$ denotes the available context budget. Let $a$ denote a candidate answer. The generation model defines a conditional generation policy over answers:
\begin{equation}
\pi_{\mathrm{model}}(a\mid q,\mathbf{X})
=
P(a\mid q,\mathbf{X}),
\end{equation}
and produces
\begin{equation}
a^{*}
=
\operatorname*{argmax}_{a}
\pi_{\mathrm{model}}(a\mid q,\mathbf{X}).
\end{equation}
where $a^{*}$ denotes the highest-probability answer under the generation policy conditioned on query $q$ and context $\mathbf{X}$.

Therefore, unlike human-facing Web search, where the ranked document list itself is the final user-facing output, AI Search requires additional processing before answer generation. The system must transform retrieved documents into an appropriate generation context so that the generation model can reliably produce high-quality answers.

This requirement reveals the gap between human-facing Search Satisfaction and answer-oriented AI Search. Once retrieved documents become inputs to a generation model, the system must make decisions beyond ranking individual documents. 
At the document level, the system must first determine whether a candidate provides information needed to generate an answer, and then whether that information is trustworthy enough to support a correct answer. At the set level, it must further organize the retained information under a finite context budget so that the generation model can produce correct answers consistently and robustly.
These requirements are not captured by standalone query--document satisfaction scores.

\begin{figure}[H]
    \centering
    \includegraphics[
        width=\linewidth
    ]{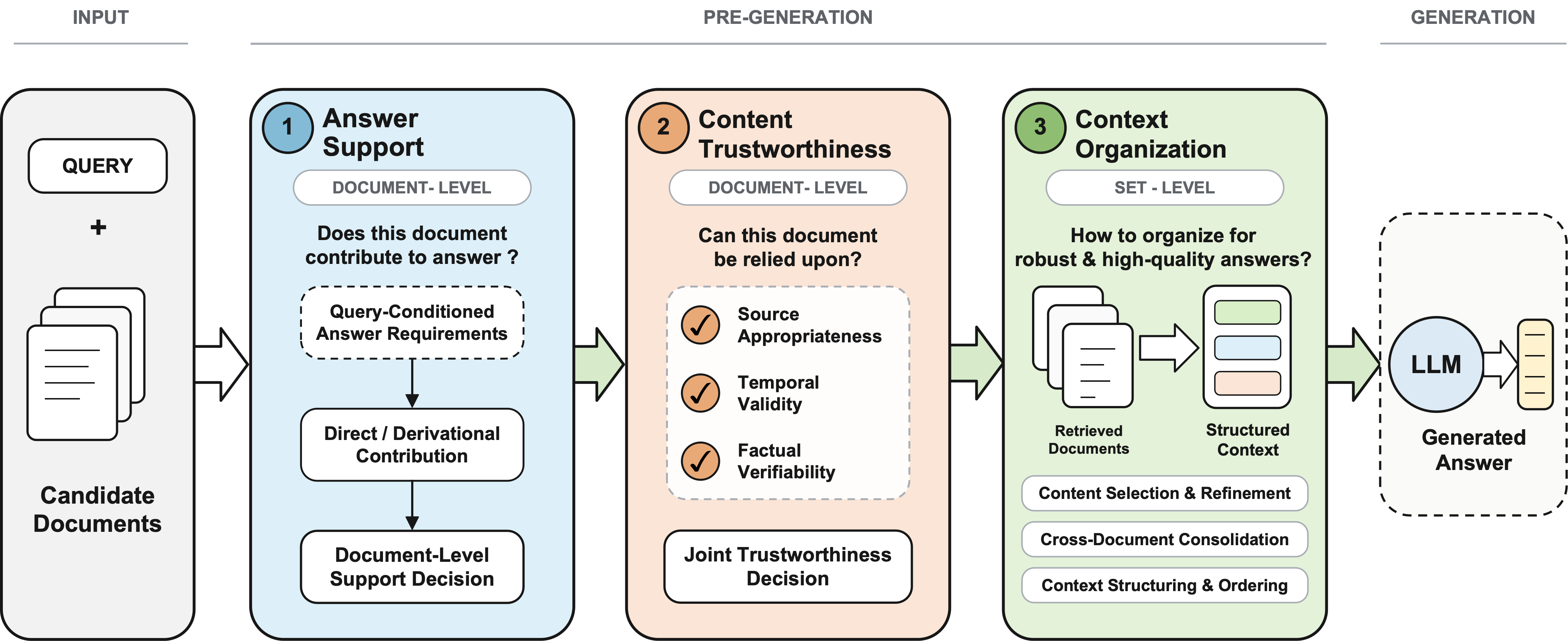}
    \caption{The proposed three-stage framework for transforming candidate documents into context for answer generation.}
    \label{fig:answer-oriented-framework}
\end{figure}

To formalize these requirements in AI search, we propose a three-stage framework for assessing retrieved information and organizing it into the context supplied to the generation model.
As illustrated in Fig.~\ref{fig:answer-oriented-framework}, the framework comprises \emph{Answer Support}, \emph{Content Trustworthiness}, and \emph{Context Organization}.

\inlinebulletsection{Stage 1: Answer Support.}
Answer Support addresses the gap between query--document relevance and actual contribution to answer construction. It is a query-conditioned, document-level criterion that evaluates whether a candidate document is answer-bearing or provides indirect evidence from which the answer can be constructed \cite{joren2024sufficient,lee2025setselection}. A document need not independently support the complete answer; it is retained when it contributes useful information to at least one informational requirement of the query \cite{zhang2026beyond}. At this stage, support refers to informational contribution rather than factual reliability.

\inlinebulletsection{Stage 2: Content Trustworthiness.}
Content Trustworthiness addresses the reliability requirements of documents consumed by the generation model. It is a document-level criterion applied to candidates retained under Answer Support and evaluates whether a document can serve as a reliable basis for accurate answer generation along three complementary dimensions. \emph{Source Trustworthiness} assesses whether the source is appropriately qualified for the content the document contributes to answer construction; \emph{Temporal Trustworthiness} assesses whether that content is applicable to the time, version, policy state, or event stage required by the query; and \emph{Information Trustworthiness} assesses whether its factual claims are verifiable and adequately substantiated by traceable evidence \cite{hwang2024rarag,jin2026sourcebench,chen2026expiration,thorne2018fever,min2023factscore}.

\inlinebulletsection{Stage 3: Context Organization.}
Context Organization addresses the shift of information selection and organization from user-side interaction to pre-generation context construction. It is a set-wise process that determines what information from the retained documents should be included in the model input and how that information should be structured and ordered under a finite context budget. The process refines irrelevant content within individual documents, consolidates overlapping information across documents, and organizes the remaining complementary information into a structured context for consistently and robustly generating correct answers~\cite{xu2023recomp,jiang2024longllmlingua,liu2024lost}.

Together, the three stages form a sequential framework that connects document-level assessment with set-level context construction.
Answer Support identifies information that contributes to answer construction, Content Trustworthiness determines whether that information can be relied upon, and Context Organization selects and organizes the retained information for consumption by the generation model.

\section{Answer Support}
\label{sec:sufficiency}

Answer Support addresses the first decision in our three-stage framework: given a query $q$ and a candidate document $d$, does $d$ provide information that contributes to the answer generation?
This contribution may be direct, through answer-bearing content, or indirect, through information needed to derive the answer. The judgment is query-conditioned and applied independently to each candidate document.

\subsection{From Query--Document Relevance to Answer Support}
\label{subsec:answer-support-relevance-limit}

In the Search Satisfaction framework, Relevance measures the human-perceived alignment between a query and a document. It captures how well the document matches the query in terms of subject, information need, and explicit constraints, considering both semantic similarity and lexical correspondence. A highly relevant document therefore closely addresses what the query asks about. However, such alignment does not necessarily indicate how much information the document contributes to the answer generation. 

\inlinebulletsection{High Query Alignment with Limited Answer Contribution.}
A document may closely match the expressed request but provide limited information that contributes to answer generation. This occurs when a document contains semantically or lexically aligned statements that appear to answer the query but provide little information beyond the stated conclusion.

For example, given the query ``What is Donnie Yen's actual height?'', a document mentioning ``I feel that Donnie Yen's real height is not even 169 cm'' is highly relevant because it matches the entity and the requested attribute, with direct lexical alignment to the query. However, the statement only provides a direct assertion without additional information that helps the generation model formulate the answer. Therefore, although the document is highly relevant to the query, the information it provides contributes little to answer generation.

\inlinebulletsection{Limited Query Alignment with High Information Contribution.}
Conversely, a document with limited direct alignment to the query may still provide information that contributes to answer generation. Such documents may not explicitly answer the query, but they can contain facts, conditions, or premises from which the answer can be derived.

For example, for a query asking how Company A's stock price may change over the next three months, a quarterly report may not directly discuss future stock-price movement and therefore receive lower Relevance. However, its description of recent financial performance, business conditions, and future plans provides information that can support the analysis of potential price trends. Such information contributes to answer generation despite the document's limited direct query alignment.

The contrast illustrates a limitation of Relevance as the sole document-level criterion for AI Search. Relevance evaluates how well a document aligns with the request expressed by the query, but such alignment does not reveal whether the document provides information that can be used to construct, derive, or explain the answer. A relevance-oriented ranker may therefore prioritize a document that directly states the requested conclusion over another document that supplies useful premises for deriving it. This motivates an additional query-conditioned, document-level criterion that explicitly evaluates a document's informational contribution to answer generation. We formalize this criterion as \emph{Answer Support} in the following subsection.

\subsection{Document-Level Answer Support}
\label{subsec:answer-support-definition}

Recent studies have shown that query--document relevance alone is insufficient for retrieval in generation-based systems, motivating new criteria that evaluate retrieved information with respect to downstream answer generation
\cite{ai2023informationretrievalmeetslarge,sun2026rethinking,lee2025setselection,trappolini-etal-2026-redefining}. Existing works can be broadly categorized into two related directions. The first direction focuses on the collective coverage or sufficiency of retrieved information. Set-level methods examine whether a collection of retrieved evidence jointly covers the information requirements of a query, while sufficient-context evaluation determines whether the resulting retrieval context contains enough information to make the query answerable \cite{lee2025setselection,joren2024sufficient}. The second direction takes an end-to-end view of retrieval utility, evaluating retrieved evidence according to its effect on downstream answer generation. In this formulation, the value of retrieved information is ultimately reflected in answer-level outcomes, such as reduced generation uncertainty or improved generation quality \cite{dai2025seper,sun2026rethinking,trappolini-etal-2026-redefining}. 
These approaches characterize retrieval quality either by the collective sufficiency of retrieved information or through its downstream effect on the generated answer.
In our framework, we instead decompose the pre-generation process into more explicit and tractable stages, each associated with a distinct assessment objective.

\begin{wrapfigure}[11]{r}{0.48\columnwidth}
    \centering
    \includegraphics[
        width=\linewidth
    ]{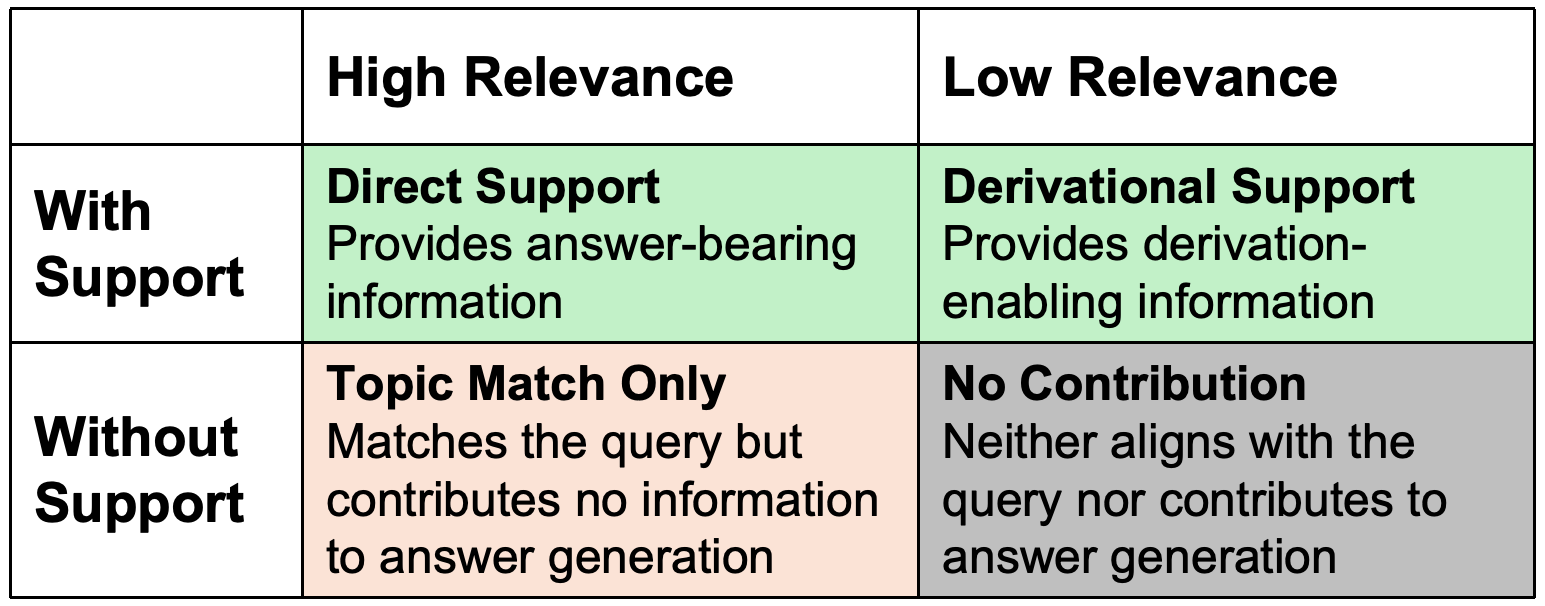}
    \caption{Relevance measures query--document alignment, whereas Answer Support measures a document's contribution to answer generation.}
    \label{fig:relevance-answer-support}
\end{wrapfigure}

At the first stage, we isolate the informational contribution of an individual candidate document from answer-level utility and set-level sufficiency, providing a document-level basis for assessing retrieved information before subsequent processing.
We define \emph{Answer Support} as a query-conditioned, document-level criterion that measures whether a candidate document provides information that contributes to answer generation. Such contribution may be direct, when the document contains answer-bearing information, or derivational, when it supplies information from which part of the answer can be derived or explained. A document need not contain all information required for answering the query; it receives credit when it contributes information to one or more parts of the answer \cite{zhang2026beyond}. As illustrated by the examples in Subsection~\ref{subsec:answer-support-relevance-limit}, query-document alignment and contribution to answer generation need not coincide. Figure~\ref{fig:relevance-answer-support} summarizes this distinction.

Formally, let
$
\mathcal{N}(q)
=
\{n_1,n_2,\ldots,n_m\}
$
denote the set of informational requirements associated with the answer scope of query $q$. Each requirement $n_i$ represents a piece of information that may contribute to the answer.
For each requirement $n_i$, the contribution of document $d$ may take either a direct or derivational form. We define the document-level Answer Support (AS) judgment as
\begin{equation}
\operatorname{AS}(q,d)
=
\operatorname{Agg}\!\left(
\left\{
\operatorname{Direct}(n_i,d)
\lor
\operatorname{Deriv}(n_i,d)
\right\}_{n_i\in\mathcal{N}(q)}
\right),
\end{equation}
where $\operatorname{Direct}(n_i,d)$ indicates that document $d$ explicitly provides answer-bearing information for requirement $n_i$, and $\operatorname{Deriv}(n_i,d)$ indicates that $d$ provides information from which the corresponding answer content can be derived or explained. $\operatorname{Agg}(\cdot)$ aggregates these requirement-level contribution judgments into the document-level Answer Support score $\operatorname{AS}(q,d)$. The output of $\operatorname{AS}(q,d)$ may be continuous or discrete depending on the specific modeling and operational requirements. Higher Answer Support indicates that the document makes a stronger contribution to the informational requirements associated with the answer scope.
Candidate documents are ranked in descending order of $\operatorname{AS}(q,d)$, yielding the Stage~1 ranked candidate set $\mathcal{C}_{\mathrm{cand}}(q)$.
Notably, Answer Support concerns informational contribution rather than the factual reliability of the contributed information.

This formulation distinguishes Answer Support from conventional Relevance. Relevance evaluates how well a document aligns with the query as expressed, whereas Answer Support evaluates whether the information contained in the document contributes to constructing the requested answer. A document therefore receives Answer Support only when it contributes to at least one answer requirement, regardless of how closely the document matches the query at the surface or semantic level. Conversely, a less directly aligned document may still provide substantial support when it supplies facts or premises for deriving part of the answer. 

In summary, Answer Support is a document-level assessment of whether a candidate document provides information that can contribute to answer generation. It operates independently for each document, leaving trustworthiness assessment and cross-document context construction to the subsequent stages of Content Trustworthiness and Context Organization.

\section{Content Trustworthiness}
\label{sec:trustworthiness}

Answer Support identifies whether a candidate document contributes to answer generation, but such contribution alone does not establish whether the document can serve as a reliable basis for generation. A document retained under the Answer Support criterion may still originate from an inappropriate source, describe a time or state that does not apply to the query, or contain factual claims that cannot be verified. We therefore define \emph{Content Trustworthiness} as the second document-level criterion in our framework: whether a document retained under Answer Support can be relied upon as a basis for answer generation \cite{ni2026trustworthyrag}.

\subsection{From Authority, Freshness, and Quality to Content Trustworthiness}
\label{subsec:trust-limitations}

In Search Satisfaction, Authority, Freshness, and Quality capture complementary conditions under which a document may constitute a high-quality result for human consumption. Authority provides a source-level prior based on the category of the source site, Freshness evaluates publication recency relative to the query's freshness demand, and Quality characterizes how clearly the document is organized and presented. These signals help users identify results that are authoritative, recent, and easy to consume, but they do not establish whether a document can serve as a reliable basis for model consumption.

The traditional Authority signal assigns an authority level according to a predefined source-site category, without conditioning the assessment on the specific query. Government websites, enterprise websites, and verified individual publishers are assigned different levels. This mechanism is useful in human-facing search, where users can inspect the website, publisher identity, and other source information before deciding whether to rely on a result \cite{rieh2002judgment,rieh2007credibility,schwarz2011augmenting}. 
However, for model consumption, the static source-category prior does not account for the alignment between the domain required by the query and the domain associated with the document source. Prior studies on source credibility have identified source expertise as a key factor in credibility assessment \cite{hilligoss2008developing}, motivating the consideration of domain alignment beyond coarse source-category grading. For example, for ``How often can ibuprofen be taken?'', both the Beijing Municipal Health Commission and a professional physician may serve as appropriate and reliable sources, although the physician may receive a lower Authority level under the source-category taxonomy.
AI Search therefore requires a query-conditioned source assessment that considers whether the source is appropriate for the domain and type of information required by the query.

The traditional Freshness signal determines whether a document is sufficiently recent by comparing its publication time with the query's freshness demand. However, publication recency does not necessarily establish whether the information itself remains temporally applicable.  
For example, a recently published article may reproduce a policy that has already been superseded, whereas an older policy document may still describe the currently effective rule if no subsequent change has taken effect. More generally, AI Search must reason about the temporal state represented by the information and whether that state matches the time required by the query \cite{zhang2021situatedqa,zhang2025mrag,cao2026re3}.
Reliability therefore requires reasoning about the validity period of the information itself and whether it has expired, changed, or been superseded.

The traditional Quality signal addresses the organization and presentation of a document for direct human consumption. A clearly structured and readable document allows users to locate and understand information more easily, but presentation quality does not establish whether the information itself is reliable. A well-organized document may still contain erroneous, poorly substantiated, or unverifiable claims. This limitation becomes more consequential in AI Search because the information is consumed by a generation model rather than directly inspected by the user, and well-presented but unreliable content may still be incorporated into an incorrect generated answer. Without explicit claim verification, unreliable information may be incorporated into the generation context and propagated into the generated answer, thereby degrading the quality of the final user-facing response. This risk motivates explicit assessment of whether the factual claims used for answer generation can be verified against qualified and traceable evidence~\cite{thorne2018fever,min2023factscore,schlichtkrull2023averitec,niu2024ragtruth,gao2023alce}

These limitations motivate a shift from human-facing Authority, Freshness, and Quality signals toward an assessment of the reliability of information used for answer generation. We capture this requirement through \emph{Content Trustworthiness}, which extends beyond static source-category authority, publication recency, and presentation quality to consider whether the information contributed by a document can serve as a reliable basis for generation. The following subsection formalizes this criterion along three complementary dimensions.

\subsection{Document-Level Content Trustworthiness}
\label{subsec:trust-definition}

Building on Answer Support defined in Subsection~\ref{subsec:answer-support-definition}, Content Trustworthiness is evaluated for documents retained as contributing to answer generation. For such a document $d$, the criterion asks whether it can serve as a reliable basis for accurate answer generation. For a query $q$ issued at time $t_0$, we define the Content Trustworthiness profile of $d$ as
\begin{equation}
\mathrm{CT}(q,d;t_0)
=
\Phi\!\left(
W_{\mathrm{src}}(q,d),
W_{\mathrm{temp}}(q,d;t_0),
W_{\mathrm{info}}(q,d)
\right),
\label{eq:content-trustworthiness}
\end{equation}
where $W_{\mathrm{src}}$, $W_{\mathrm{temp}}$, and $W_{\mathrm{info}}$ denote Source, Temporal, and Information Trustworthiness, respectively, and $\Phi(\cdot)$ maps these dimension-specific outputs into a unified document-level Content Trustworthiness judgment. All three components are document-level assessments, characterizing complementary aspects of reliability: whether the source of $d$ is appropriate, whether its contributed content is temporally applicable, and whether its factual claims are verifiable. Across these three dimensions, the assessment draws on external knowledge and evidence beyond document-local signals, including source-side background knowledge, temporal-state information, and traceable evidence for factual verification \cite{jin2026sourcebench,zhang2021situatedqa,schlichtkrull2023averitec}.
 
\inlinebulletsection{Source Trustworthiness.}
Motivated by prior work on source reliability and source-specific authority \cite{hwang2024rarag,jin2026sourcebench,morlan2026factappeal}, we extend the source assessment beyond the static source-category grading represented by the traditional Authority signal $A(q,d)$ in Eq.~\ref{eq:authority}. Source Trustworthiness is not equivalent to this Authority assessment:
$$
W_{\mathrm{src}}(q,d)
\not\equiv
A(q,d).
$$

Let $p_d$ denote the content producer of document $d$. We characterize Source Trustworthiness through two complementary assessment paths. Sources that receive a sufficiently high conventional Authority level under the predefined source-category taxonomy, such as government or enterprise websites,  directly satisfy the source requirement. For sources with lower traditional Authority, we further consider \emph{Domain Alignment} and the \emph{Producer Profile}. Conceptually,
\begin{equation}
W_{\mathrm{src}}(q,d)
=
\Phi_{\mathrm{src}}\!\left(
A(q,d),
\operatorname{Domain}(q,d)
\land
\operatorname{P}(p_d)
\right).
\label{eq:source-trustworthiness}
\end{equation}
Here, $A(q,d)$ denotes the conventional Authority signal defined in Eq.~\ref{eq:authority}. \emph{Domain Alignment}, denoted by $\operatorname{Domain}(q,d)$, represents the alignment category between the domain required by query $q$ and the domain associated with document $d$. It allows domain-specialized sources that may receive a lower traditional Authority level to remain eligible when their domain is well aligned with the query. \emph{Producer Profile}, denoted by $\operatorname{P}(p_d)$, represents the profile level of content producer $p_d$ based on three groups of signals: identity and qualification signals, such as identity, verification status, and domain expertise; historical credibility signals, such as reputation, publication history, and representative content; and behavioral and audience signals, such as audience scale and engagement, publishing frequency, and behavioral patterns.$\Phi_{\mathrm{src}}(\cdot)$ maps these heterogeneous source-side signals into a unified Source Trustworthiness judgment $W_{\mathrm{src}}(q,d)$. The resulting $W_{\mathrm{src}}(q,d)$ may be represented as either a discrete level or a continuous score depending on the specific modeling and operational setting.

Source Trustworthiness thus extends Authority from static source-category grading to query-conditioned, producer-aware suitability of the source for the information used.

\inlinebulletsection{Temporal Trustworthiness.}
Temporal Trustworthiness extends the traditional Freshness signal $T(q,d;t_0)$ defined in Eq.~\ref{eq:freshness}. While Freshness primarily models the compatibility between the publication time of document $d$ and the freshness demand of query $q$, Temporal Trustworthiness evaluates whether the temporal state represented by the document remains applicable to the query at search time $t_0$ \cite{zhang2021situatedqa,cao2026re3}. Accordingly,
$$
W_{\mathrm{temp}}(q,d;t_0)
\not\equiv
T(q,d;t_0).
$$

More specifically, let $\tau_{\mathrm{cont}}(d)$ denote the content time of document $d$, i.e., the time associated with the state or information described in the document, and let $\tau_{\mathrm{exp}}(q,d;t_0)$ denote the query- and search-time-conditioned expiration time beyond which that information is no longer considered applicable. Temporal Trustworthiness then evaluates whether the document remains temporally applicable at the current search time:
\begin{equation}
W_{\mathrm{temp}}(q,d;t_0)
=
\Phi_{\mathrm{temp}}\!\left(
\tau_{\mathrm{cont}}(d)
-
\tau_{\mathrm{exp}}(q,d;t_0)
\right),
\label{eq:temporal-trustworthiness}
\end{equation}
where $\Phi_{\mathrm{temp}}(\cdot)$ maps the relative temporal position of the document content time with respect to the query-conditioned expiration time into a Temporal Trustworthiness judgment. Specifically, the difference $\tau_{\mathrm{cont}}(d)-\tau_{\mathrm{exp}}(q,d;t_0)$ indicates whether, and by how much, the content time falls before or after the expiration boundary. The resulting $W_{\mathrm{temp}}(q,d;t_0)$ may be represented as either a discrete level or a continuous score depending on the specific modeling and operational setting.

Temporal Trustworthiness therefore goes beyond publication recency by reasoning about both the temporal state represented by the document and whether the information remains valid at the current search time.

\inlinebulletsection{Information Trustworthiness.}
Information Trustworthiness extends beyond the Quality signal $Q(q,d)$ defined in Eq.~\ref{eq:quality}. While Quality evaluates the organization and presentation of a document for direct human consumption, Information Trustworthiness evaluates the factual reliability of the claims in $d$ that contribute to answer generation. Accordingly,
$$
W_{\mathrm{info}}(q,d)
\not\equiv
Q(q,d).
$$

Let
$
\mathcal{F}(q,d)
=
\{c_1,c_2,\ldots,c_n\}
$
denote the set of externally verifiable factual claims in document $d$ that contribute to answer generation. For each claim $c\in\mathcal{F}(q,d)$, let $\mathcal{E}_c$ denote the qualified and traceable evidence available for verification. The evidence-relative verification status of $c$ is
$
V(c\mid\mathcal{E}_c)
\in
\left\{
\mathrm{corroborated},
\mathrm{contradicted},
\mathrm{unresolved}
\right\},
$
where \emph{corroborated} indicates agreement with the available evidence, \emph{contradicted} indicates conflict with the available evidence, and \emph{unresolved} indicates that the evidence is insufficient or conflicting to determine either outcome \cite{thorne2018fever,min2023factscore,schlichtkrull2023averitec}.

For documents containing externally verifiable factual claims, the claim-level verification outcomes are aggregated into document-level Information Trustworthiness as
\begin{equation}
W_{\mathrm{info}}(q,d)
=
\Phi_{\mathrm{info}}\!\left(
\left\{
V(c\mid\mathcal{E}_c)
\right\}_{c\in\mathcal{F}(q,d)}
\right),
\label{eq:info-trustworthiness}
\end{equation}
where $\Phi_{\mathrm{info}}(\cdot)$ maps the claim-level verification outcomes into a unified document-level Information Trustworthiness judgment. The mapping follows a non-compensatory principle: unresolved or contradicted factual claims should not be offset by corroborated claims elsewhere in the document. Depending on the specific modeling and operational setting, $W_{\mathrm{info}}(q,d)$ may be represented as either a discrete level or a continuous score.

Information Trustworthiness therefore shifts the assessment from document presentation to evidence-based verification of the factual claims that contribute to answer generation.

\par\smallskip
\noindent\textbf{Joint Trustworthiness Decision.}
Under the joint Content Trustworthiness formulation in Eq.~\ref{eq:content-trustworthiness}, the three complementary dimensions are mapped into an overall score that can be used to further rank or filter $\mathcal{C}_{\mathrm{cand}}(q)$, yielding $\mathcal{C}_{\mathrm{ret}}(q)$. The assessment is non-compensatory: the weakest applicable dimension acts as the limiting factor, so a serious deficiency in source suitability, temporal applicability, or factual reliability should not be offset by stronger assessments in the others.

In the final decision, the dimension that becomes the limiting factor may vary with the query and the role of the document in answer generation. For a factoid query, retained documents often contain factual claims that directly determine the answer, making Information Trustworthiness more likely to become the limiting dimension when such claims are contradicted or unresolved. For an experience-oriented query, Source Trustworthiness or Temporal Trustworthiness may instead become the limiting dimension, depending on whether the producer is an appropriate source of first-hand experience and whether that experience remains temporally applicable. Because subjective observations are not themselves externally verifiable factual claims, Information Trustworthiness applies only to any verifiable factual claims contained in or underlying such content; when no such claims are involved, it is treated as not applicable and does not constrain the joint assessment.

\section{Context Organization}
\label{sec:organization}

The first two stages operate at the document level, assessing candidate documents in terms of their contribution to answer generation and their reliability as a basis for generation. Documents that satisfy these assessments form the retained set $\mathcal{C}_{\mathrm{ret}}(q)$, which serves as the input to Context Organization. The remaining problem is to construct the generation context supplied to the generation model: what content should be selected from the retained documents, and how should that content be structured and ordered under a finite context budget? We define \emph{Context Organization} to address this problem.

Naively concatenating the retained documents does not necessarily produce an effective generation context. Individual documents may contain content that does not contribute to the requested answer, and the retained set may contain redundancy, and complementary information may be distributed across multiple sources.
Under a finite context budget, such content can crowd out more useful information, and make the information needed for answer construction harder for the generation model to identify and use. Context Organization therefore operates both within and across retained documents, determining \emph{what content and associated metadata to include} and \emph{how the selected content should be structured and ordered} in the context supplied to the generation model.

\subsection{From Ranked Lists to Generation Context}
\label{subsec:organization-limitations}

In traditional Web search, retrieved documents are scored according to Search Satisfaction and presented to users as a ranked list. This representation is well suited to human consumption: users can decide which results to inspect, identify useful information within individual pages, compare information across documents, and synthesize the needed information on their own. The search system therefore primarily determines the ordering of documents, while information selection and cross-document synthesis are largely left to the user.

However, this ranked-list representation does not satisfy the requirements of model-facing context construction in AI Search. Unlike human users, who can selectively inspect and synthesize information across ranked results, a generation model receives a bounded context and therefore depends on the system to determine what information should be included and how it should be organized. These decisions remain necessary even after document-level assessment: the retained documents may collectively contain the information required for a correct answer, yet individual documents can still contain irrelevant content, multiple documents can provide redundant or overlapping information, and useful evidence can be distributed or poorly positioned within a long context \cite{yoran2023irrelevant,liu2024lost}. These properties can consume the finite context budget and make the information needed for answer construction more difficult for the generation model to identify and use.

AI Search must therefore move information selection and organization from user-side interaction to pre-generation context construction. Given the documents retained after document-level assessment, the system must determine what information and associated metadata should enter the model context, how overlapping or complementary information across documents should be handled, and how the resulting content should be structured and ordered under a finite context budget. This requires both removing unnecessary content within individual documents and jointly organizing information across the retained document set \cite{qian2024cfic,yoon2024compact,jin2025longrefiner,lee2025setselection,sarthi2024raptor}.

We define \emph{Context Organization} as this set-wise context-construction process. Rather than directly passing a ranked list of documents to the generation model, Context Organization selects and organizes information from the retained documents into the context supplied for generation, and preserves the metadata and contextual relations needed for the selected information to be interpreted correctly.

\subsection{Set-Level Context Organization}
\label{subsec:organization-definition}

Let $\mathcal{C}_{\mathrm{ret}}(q)$ denote the set of documents retained after the document-level Answer Support and Content Trustworthiness assessments. Following the formulation in Eq.~\ref{eq:context-organization}, Context Organization jointly operates over $\mathcal{C}_{\mathrm{ret}}(q)$ under a finite context budget $B$ to construct the generation context $\mathbf{X}$. Unlike the preceding stages, which assess candidate documents independently, Context Organization considers the retained document set jointly because the inclusion, representation, and placement of information from one document may depend on information available in others. Its objective is not to shorten the input indiscriminately, but to construct a bounded context that preserves and organizes the information needed for the generation model to consistently and robustly produce correct answers.

\begin{figure}[H]
    \centering
    \includegraphics[
        width=\linewidth
    ]{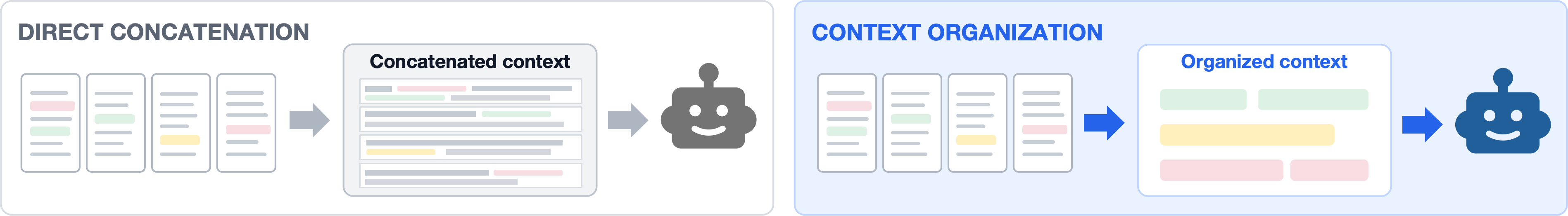}
    \caption{
    Comparison between direct document concatenation and Context Organization. Context Organization filters non-contributing and redundant content and reorganizes complementary information by answer aspect, producing a structured generation context that facilitates consistent and robust generation of correct answers.
    }
    \label{fig:organization-case}
\end{figure}

Context Organization makes two coupled decisions: \emph{what information to include} and \emph{how the selected information should be organized}. The first determines which content from the retained documents should enter the generation context. Content that does not contribute to answer construction or that duplicates information available elsewhere may be removed, while complementary information and the qualifiers or metadata needed to interpret and attribute the retained content should be preserved \cite{qian2024cfic,yoon2024compact,jin2025longrefiner}. The second decision determines how the selected content should be structured and ordered across documents. Information associated with different entities, versions, temporal states, or sources should remain distinguishable where these distinctions affect interpretation, and complementary content should be organized so that the generation model can identify and use it effectively under the finite context budget \cite{lee2025setselection,sarthi2024raptor,liu2024lost}. Figure~\ref{fig:organization-case} illustrates the distinction between direct document concatenation and Context Organization: rather than passing retained documents to the generation model as a flat sequence, Context Organization selects and structures the retained information into an organized context before generation.

The objective of Context Organization is to construct a context that enables the generation model to reliably produce high-quality answers from the retained information. This objective must be achieved without distorting or omitting information required for answer generation. For the context organization process $O$ defined in Eq.~\ref{eq:context-organization}, we characterize its admissible output through two constraints: \emph{Faithfulness} and \emph{Richness}.

\begin{itemize}
    \item \emph{Faithfulness.}
    The organized context $\mathbf{X}$ should preserve the meaning, attribution, and qualifying conditions of the information in $\mathcal{C}_{\mathrm{ret}}(q)$ during selection, compression, consolidation, and restructuring, without introducing unwarranted information.
    
    \item \emph{Richness.}
    The organized context $\mathbf{X}$ should preserve sufficient informational breadth and depth for answer generation. Breadth requires retaining the range of information needed for the requested answer, while depth requires preserving adequate detail, conditions, and explanatory information for the retained aspects. Together, they prevent $O$ from omitting necessary information or reducing it to shallow representations.
\end{itemize}

Together, these constraints define the admissible transformation space for Context Organization. The organization process may remove unnecessary content, consolidate overlapping information, and restructure the retained content under the context budget, but it must preserve semantic fidelity, required informational breadth, and sufficient depth. Context Organization is therefore a constrained context-construction process rather than unconstrained compression or summarization.

\section{Model Implementation and Optimization}
\label{sec:model-optimization}

The preceding sections define three pre-generation objectives for assessing retrieved information and constructing the context supplied to the generation model: \emph{Answer Support}, \emph{Content Trustworthiness}, and \emph{Context Organization}. This section describes how these objectives are operationalized and optimized in industrial practice. 

We distinguish two complementary optimization paradigms according to when the optimization signal becomes available relative to answer generation. \emph{Prior optimization} operates before generation and improves the retrieval and context-construction decisions defined by the three stages. \emph{Posterior optimization} operates after generation, using answer-level outcomes to diagnose and trace deficiencies back to earlier retrieval and context-construction decisions.

\subsection{Prior Optimization in Practice}
\label{subsec:prior-optimization}

Prior optimization operationalizes the three-stage framework before answer generation, with the goal of enabling the generation model to consistently and robustly produce correct answers based on the supplied context. The following describes how Answer Support, Content Trustworthiness, and Context Organization are implemented in practice.

\subsubsection{LLM-based Answer Support Document Ranker}
\label{subsec:support-practice}

Answer Support ranking requires a different capability from conventional relevance ranking. A relevance model primarily estimates whether a document matches the subject, intent, and explicit constraints of a query. In contrast, an Answer Support ranker must determine whether the document contains information that can materially contribute to answering the query, either by directly providing part of the answer or by supplying facts and premises from which the answer can be derived. The latter judgment can require broader contextual knowledge, implicit constraint resolution, and multi-step reasoning, particularly when useful documents do not directly restate the requested conclusion. 

Considering the strict latency constraint, existing industrial online rankers are typically built on relatively small models (e.g., BERT\cite{devlin2019bert}), and are optimized primarily for query--document relevance. Their limited model capacity and short input length are sufficient for many direct semantic-matching cases but constrain their ability to reason over long documents and assess indirect information contributions. We therefore upgrade the ranker to a larger language-model backbone and formulate it as a reasoning-oriented document ranker. The larger backbone provides greater capacity for understanding complex queries, resolving implicit relations, and determining whether a document can contribute to answer generation even when the query and the supporting content have limited surface lexical overlap.

\begin{figure}[H]
    \centering
    \includegraphics[
        width=\linewidth
    ]{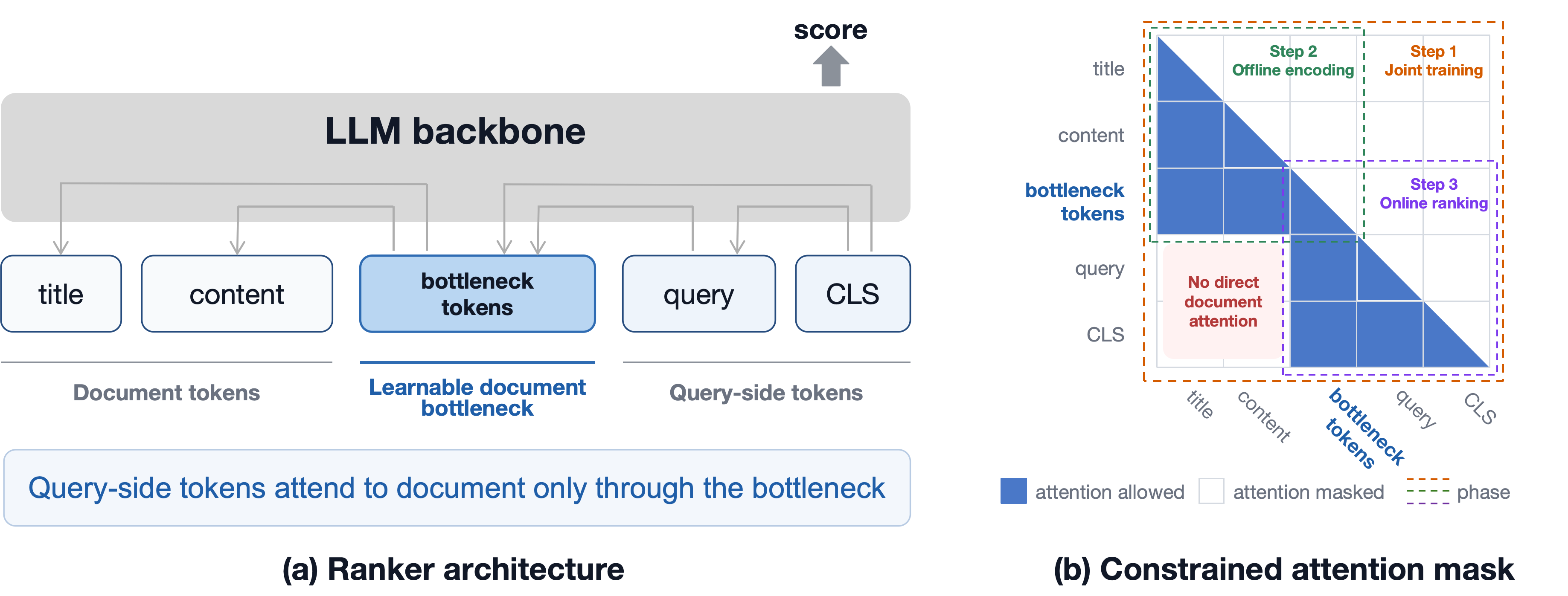}
    \caption{
    Architecture and constrained attention design of the LLM-Based Document Ranker.
    (a) Document information is compressed into learnable bottleneck tokens, which mediate query-side access for scoring.
    (b) The constrained attention mask enables joint training, offline document encoding, and online query-conditioned ranking without direct query-to-document attention.
    }
    \label{fig:answer-support-model}
\end{figure}

However, directly applying a larger backbone to full documents during online serving would incur prohibitive computation and latency. To address this issue, inspired by \cite{mu2023gist}, which compresses static, reusable prompts into gist-token representations that can be cached for efficient inference, we introduce a compression-based LLM ranker. As illustrated in Figure~\ref{fig:answer-support-model}, the model compresses each document into a small number of learnable bottleneck tokens and restricts query-side access to the document through this compressed representation. 
During training, document encoding and query-conditioned ranking are optimized jointly under a single constrained attention mask. The constrained attention mask allows the bottleneck tokens to aggregate information from the document, while preventing query-side tokens from attending directly to the original document tokens. Consequently, a document with $L_d$ tokens is compressed into $M$ bottleneck representations, and $M \ll L_d$. That is,
\begin{equation}
\mathbf{g}(d)
=
\operatorname{Compress}_{\phi}
\left(
\operatorname{title}(d),
\operatorname{content}(d)
\right)
=
\left[
\mathbf{g}_1(d),\ldots,\mathbf{g}_M(d)
\right],
\qquad
M\ll L_d,
\label{eq:gist}
\end{equation}
where $\mathbf{g}_1(d),\ldots,\mathbf{g}_M(d)$ denote the hidden states of the $M$ document-bottleneck tokens.

To align the ranker with the Answer Support objective, the supervision target is changed from relevance to answer contribution. Positive examples include documents that directly provide answer-bearing information as well as those that supply necessary facts, conditions, or derivational premises, whereas hard negatives are topically or semantically aligned documents that do not contribute useful information for answering the query. Training with these labels directs the larger backbone toward distinguishing substantive answer contribution from superficial query--document alignment.

At inference time, the same model can be decomposed into offline document encoding and online query-conditioned ranking. Because $\mathbf{g}(d)$ is query-independent, the bottleneck representation and its reusable key--value states can be computed and cached offline. When a query arrives, the online ranker therefore operates only on the compact bottleneck representation rather than the full document sequence, accessing document information through these cached states to produce an estimated Answer Support score:
\begin{equation}
\widehat{\mathrm{AS}}(q,d)
=
\operatorname{Rank}_{\theta}
\left(
q,
\mathbf{g}(d)
\right).
\end{equation}

This design substantially reduces the amount of document information that must participate in online ranking, replacing the full document sequence with a compact set of bottleneck tokens. It therefore allows the ranker to exploit information from substantially longer documents while keeping the online computation tractable under serving-time latency constraints.
The resulting ranker assigns each candidate document a continuous estimated Answer Support score $\widehat{\mathrm{AS}}(q,d)$ and ranks the candidates accordingly. The ranked candidates are then further assessed by Content Trustworthiness, after which the retained documents form $\mathcal{C}_{\mathrm{ret}}(q)$ and are passed to set-level Context Organization.

\subsubsection{Document Trustworthiness Degree Assessment}
\label{subsec:trust-practice}

\begin{figure}[H]
    \centering
    \includegraphics[
        width=\linewidth
    ]{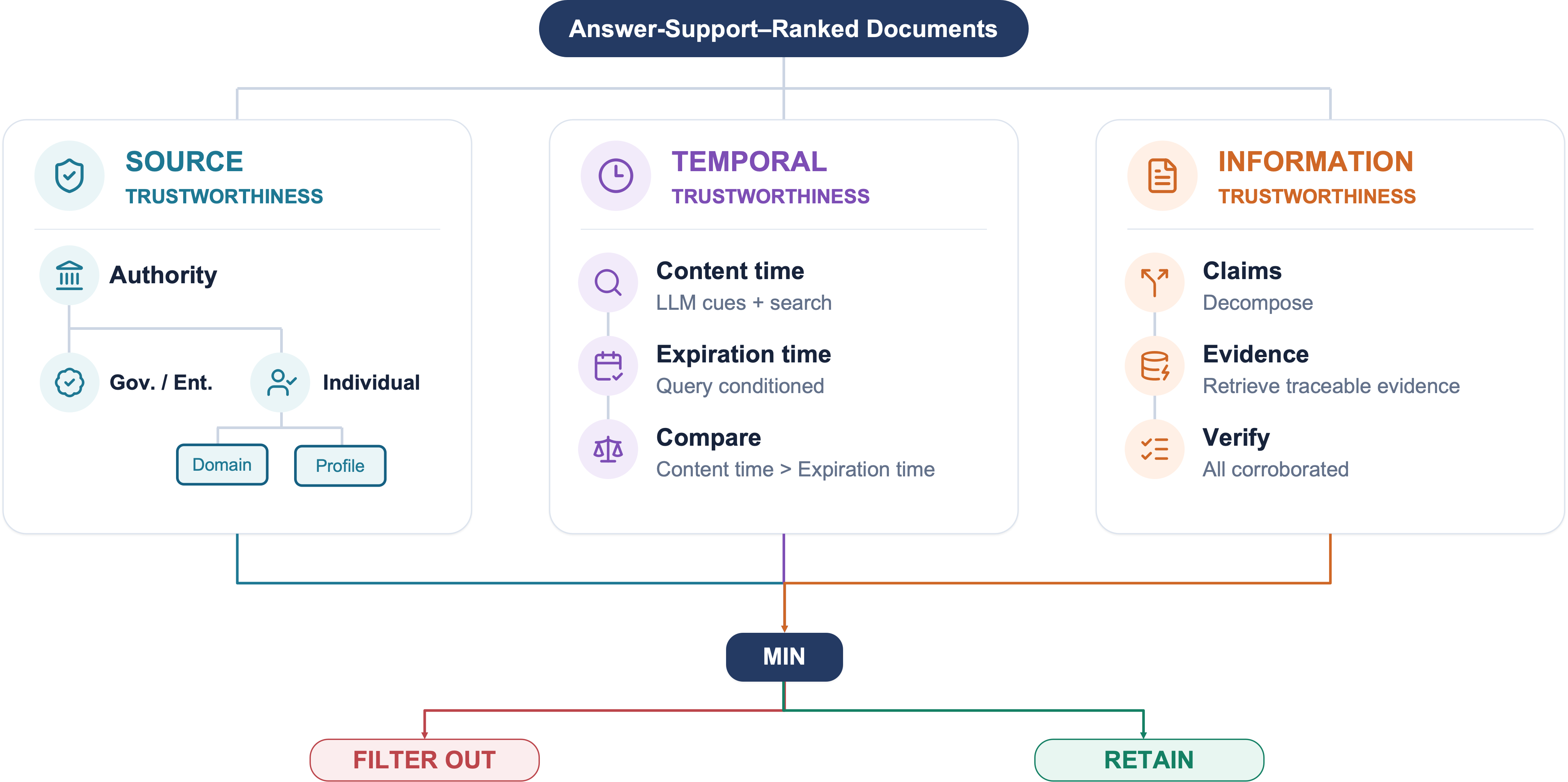}
    \caption{
    Industrial framework for document-level Content Trustworthiness assessment.
    Given Answer-Support-ranked documents, the system evaluates source suitability, temporal validity, and factual reliability, respectively,
    and then combines the three judgments with a minimum-based rule to decide
    whether each document is retained or filtered out.
    }
    \label{fig:trust-framework}
\end{figure}

Content Trustworthiness further evaluates the candidate documents ranked by their Answer Support scores along three dimensions: \emph{Source Trustworthiness}, \emph{Temporal Trustworthiness}, and \emph{Information Trustworthiness}. In current industrial practice, these dimensions are operationalized through dedicated capabilities for source assessment, temporal-validity modeling, and claim-level verification. Figure~\ref{fig:trust-framework} illustrates the overall industrial framework for document-level Content Trustworthiness assessment.

\inlinebulletsection{Source Trustworthiness.}
The assessment first applies the conventional Authority grading defined by $A(q,d)$. Under the source taxonomy, government websites receive Authority level~3, enterprise websites level~2, verified individual sources level~1, and ordinary individual sources level~0. Government and enterprise sources are treated as sufficiently reliable at this stage and can directly pass the Source Trustworthiness assessment.
Individual sources, including verified individual sources and ordinary individual sources, are further assessed through \emph{Domain Alignment} and \emph{Producer Profile}. For Domain Alignment, the query is classified under a predefined taxonomy of 29 domains, and $\operatorname{Domain}(q,d)$ characterizes the correspondence between the domain required by query $q$ and the domain associated with document $d$. The alignment is categorized as \textsc{Full}, \textsc{Partial}, or \textsc{Absent}. In the current implementation, only fully domain-aligned individual sources remain eligible for further source verification.
Producer Profile, denoted by $\operatorname{P}(p_d)\in\{0,1\}$, evaluates whether the content producer exhibits a reliable profile. The assessment incorporates signals such as identity and verification status, professional credentials, publication history, representative content, audience scale and engagement, publishing frequency, and behavioral patterns. A value of $1$ indicates that the producer profile passes the reliability check, whereas $0$ indicates profile-level abnormalities or insufficient evidence of a reliable producer. An individual source passes only when it is fully aligned with the query domain and its Producer Profile verification succeeds.

Following the source-assessment logic in Eq.~\ref{eq:source-trustworthiness}, the final $W_{\mathrm{src}}(q,d)$ is binary. Government and enterprise sources that satisfy the conventional Authority requirement receive $W_{\mathrm{src}}(q,d)=1$ directly, and individual sources receive $W_{\mathrm{src}}(q,d)=1$ only when $\operatorname{Domain}(q,d)=\textsc{Full}$ and $\operatorname{P}(p_d)=1$. When the query explicitly requires an official source, the assessment applies a hard constraint under which only official sources are eligible, and an official source receives $W_{\mathrm{src}}(q,d)=1$ only if $A(q,d)=1$.

\inlinebulletsection{Temporal Trustworthiness.}
Temporal assessment distinguishes when a document is published from when its information actually applies \cite{zhang2021situatedqa,cao2026re3}. A recently published page may reproduce an earlier event state, an expired policy, or an obsolete research finding. Publication time alone therefore misclassifies republished content as current and allows outdated information to affect generation \cite{ouyang2025hoh}. Temporal Trustworthiness consequently requires two complementary capabilities: identifying the time represented by the content and determining whether that content remains valid for the current query.

\par\medskip
\noindent\emph{Content-time identification.}
This component identifies the temporal state represented by the information contributing to answer generation, including the relevant time period, effective interval, version, policy state, or event stage. It provides the temporal representation required to determine whether the content remains applicable to the query.

The implemented temporal-localization module predicts a normalized day-level estimate of the document's content time when explicit metadata are missing, unreliable, or conflicting. Given the document body, an LLM agent analyzes available temporal cues and selects among reasoning, search, and answer actions. When explicit evidence exists, the agent extracts and normalizes the corresponding cues; otherwise, it performs adaptive retrieval to acquire additional evidence from search results, including titles, snippets, page content, and publication dates. The agent iteratively updates its temporal estimate within a bounded search budget.

The module distinguishes content time from other temporal signals, such as event dates, update dates, quoted-source dates, historical references, and reposting traces. Search-derived publication dates are treated as auxiliary evidence rather than direct prediction targets. The final estimate integrates temporal cues from the document and evidence acquired through search. The search-and-reasoning policy is optimized with an actor--critic PPO objective using temporally shaped rewards, including date-distance rewards and search-behavior rewards that balance unnecessary retrieval avoidance and evidence acquisition.

The current implementation provides day-level content-time localization but does not yet recover complete effective intervals, version lineages, policy states, or event stages.

\par\medskip
\noindent\emph{Expiration-time identification.}
Given the identified temporal state, the system estimates a query-conditioned expiration boundary that separates temporally applicable content from potentially stale content at the current search time \cite{chen2026expiration}. The implemented module first identifies queries sensitive to stale information. For these queries, it collects highly ranked documents, publication times, and source-authority signals, which are then used by an LLM to predict a query-specific expiration cutoff. Uncertain predictions are further examined through an agentic reflection process and, when unresolved, routed to human review before activation.

Validated expiration cutoffs are stored in a query-keyed dictionary. During serving, matched queries retrieve their corresponding cutoffs, and candidate documents published before the boundary are treated as potentially expired. Among documents satisfying the required Answer Support assessment, temporally admissible documents are promoted while preserving their original relative order.

The current implementation operationalizes temporal validity through query-level expiration boundaries and document publication times. It therefore provides an approximation for identifying potentially expired results rather than a complete Temporal Trustworthiness decision. The final temporal assessment jointly considers the expiration boundary and the content time identified above.

\par\medskip
\noindent\emph{Joint temporal decision.}
The two components are combined according to the Temporal Trustworthiness criterion defined in Eq.~\ref{eq:temporal-trustworthiness}. The content-time module provides $\tau_{\mathrm{cont}}(d)$, and the expiration-time module provides the query-conditioned boundary $\tau_{\mathrm{exp}}(q,d;t_0)$. The mapping function $\Phi_{\mathrm{temp}}(\cdot)$ evaluates the relative position between the content time and query-conditioned expiration boundary. In the current implementation, it produces a binary mapping: $W_{\mathrm{temp}}(q,d;t_0)=1$ when $\tau_{\mathrm{cont}}(d)-\tau_{\mathrm{exp}}(q,d;t_0)>0$, indicating temporal applicability, and $W_{\mathrm{temp}}(q,d;t_0)=0$ when the difference is below $0$, indicating temporal expiration.

\inlinebulletsection{Information Trustworthiness.}
The assessment first decomposes the factual content in each candidate document into independently verifiable claims. For each claim, the system formulates a verification query and retrieves external evidence with explicit provenance.

The qualified evidence is then compared with the claim to determine its verification state. Following the claim-level verification framework defined in Section~\ref{subsec:trust-definition}, the industrial verifier assigns each claim one of three statuses: \emph{corroborated}, \emph{contradicted}, or \emph{unresolved}. A claim is \emph{corroborated} when qualified evidence agrees with it, \emph{contradicted} when qualified evidence conflicts with it, and \emph{unresolved} when the available evidence is insufficient to determine either conclusion. In particular, an unresolved claim is not treated as correct because no contradiction has been identified.

The current implementation operationalizes this verification process through document-level factual-risk screening. For each candidate document, the system decomposes the factual content contributing to answer generation into independently verifiable claims. A retrieval query is generated for each claim, and the claim is checked against the retrieved evidence to determine whether it can be corroborated. The resulting evidence is then used to assign one of the three verification statuses defined above: \emph{corroborated}, \emph{contradicted}, or \emph{unresolved}. 

Following the aggregation rule defined in Eq.~\ref{eq:info-trustworthiness}, the claim-level verification outcomes are combined into the document-level Information Trustworthiness assessment. In the current implementation, this assessment is operationalized as a binary signal: $W_{\mathrm{info}}(q,d)=1$ only when all externally verifiable claims that contribute to answer generation are corroborated; if any such claim is contradicted or unresolved, $W_{\mathrm{info}}(q,d)=0$. Thus, both verified factual inconsistency and insufficient verification evidence prevent the document from being treated as factually reliable.

\medskip
\noindent\textbf{Integration across trustworthiness dimensions.}
The general joint formulation in Eq.~\ref{eq:content-trustworthiness} is instantiated in our industrial implementation as a binary, non-compensatory decision. Specifically, Source Trustworthiness, Temporal Trustworthiness, and Information Trustworthiness are each mapped to $\{0,1\}$, where $1$ indicates that the document satisfies the corresponding criterion and $0$ otherwise. 
In this implementation, the mapping function $\Phi(\cdot)$ in Eq.~\ref{eq:content-trustworthiness} is instantiated as a minimum aggregation over the three dimensions.
Thus, a document is retained only when it passes all three applicable trustworthiness assessments. Documents with $W_{\mathrm{CT}}(q,d)=0$ are filtered out, while those with $\mathrm{CT}(q,d)=1$ preserve their Stage~1 Answer Support ordering and collectively form $\mathcal{C}_{\mathrm{ret}}(q)$. The resulting retained set is then passed to Stage~3 Context Organization for set-level processing.

\subsubsection{Set-wise Document Selector and Context Organization}
\label{subsec:organization-practice}

After the document-level Answer Support and Content Trustworthiness assessments, Context Organization operates jointly over the retained documents in $\mathcal{C}_{\mathrm{ret}}(q)$ to construct the input context for the generation model. It addresses two connected decisions: what information should be selected and refined from the retained documents, and how that information should be consolidated, structured, and ordered across documents. The objective is to construct a bounded context that enables the generation model to consistently and robustly produce correct answers.
These two decisions can be implemented either sequentially or jointly. In the \emph{pipeline-based architecture}, within-document refinement and cross-document organization are performed as separate successive steps. An \emph{Extractive Extractor} or \emph{Generative Extractor} first refines each candidate document, after which a \emph{Set-wise Organizer} consolidates and arranges the retained content across documents. By contrast, the \emph{end-to-end architecture} for Context Organization integrates generative extraction and set-wise organization within a single \emph{Unified Extractor--Organizer}, jointly optimizing within-document refinement and cross-document organization under a shared objective.

\begin{figure}[H]
    \centering
    \includegraphics[
        width=\linewidth
    ]{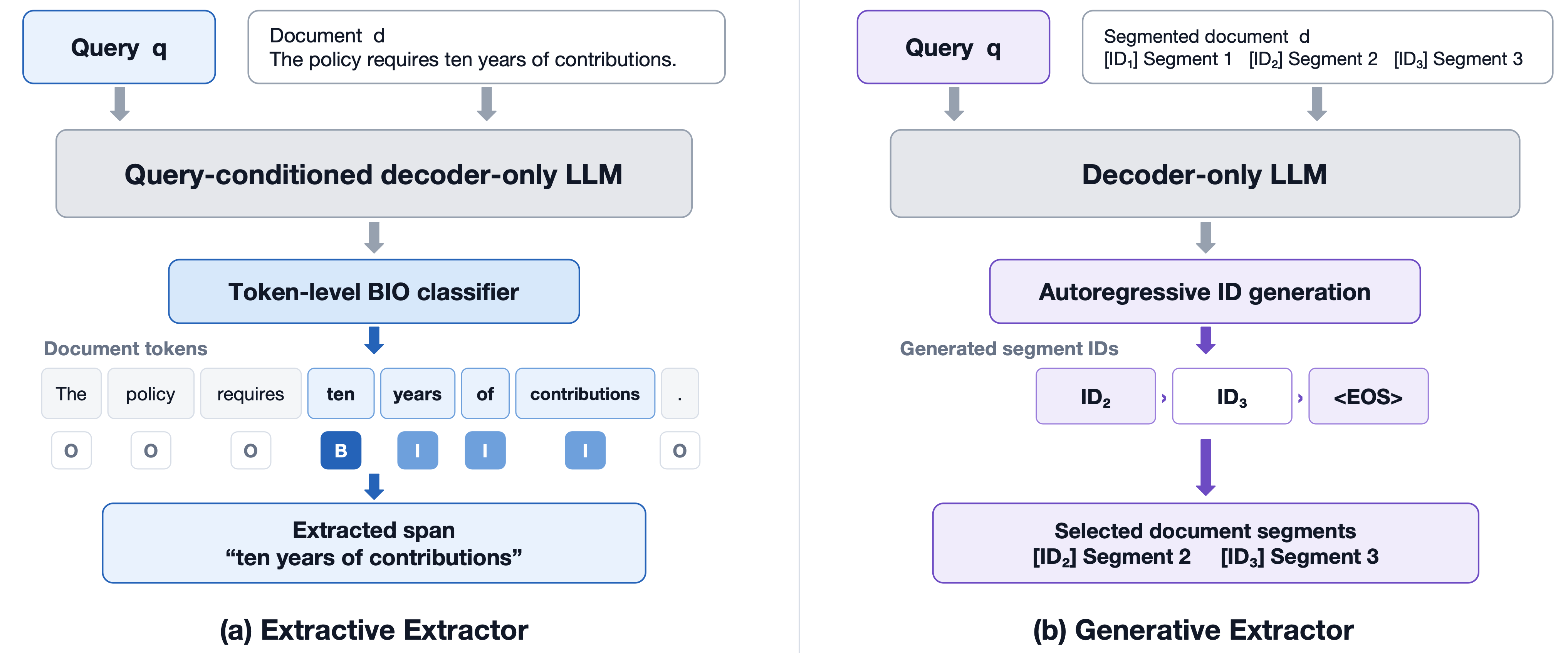}
    \caption{
    Two document-level extraction modules for Context Organization. 
    (a) The \emph{Extractive Extractor} uses query-conditioned BIO tagging to identify answer-contributing spans while preserving their original wording. 
    (b) The \emph{Generative Extractor} autoregressively selects indexed document segments for context construction.
    }
    \label{fig:extractor-model}
\end{figure}

\inlinebulletsection{Within-document refinement.}
The retained documents supplied to this stage have already undergone page parsing, with common structural noise such as navigation elements, advertisements, and page templates removed. As illustrated in Fig.~\ref{fig:extractor-model}, an \emph{Extractive Extractor} identifies spans that contribute to answer generation and preserves their original wording and local context. When the required information is distributed across distant passages in a long document or encoded through structural relations, e.g., a complex table, a \emph{Generative Extractor} consolidates the contributing information into a compact and coherent representation. Together, these two extraction modes determine what information should be preserved from each document for subsequent set-wise processing.

\inlinebulletsection{Cross-document consolidation and organization.}
The \emph{Set-wise Organizer} jointly processes the content units produced by within-document refinement across multiple retained documents. It identifies units that convey the same information under equivalent applicability conditions and consolidates them into a shared representation while preserving their provenance. Content is kept separate when it contributes complementary information, introduces a distinct applicability condition, or provides distinct provenance or additional evidence. The resulting units are then grouped according to the informational requirements of the query. Distinctions among entities, temporal states, versions, and sources are preserved whenever merging them would alter the meaning, applicability, or provenance of the content.

After consolidation and grouping, the \emph{Set-wise Organizer} constructs the final context in three steps. First, it produces a concise summary for each information cluster while retaining the underlying content units. Second, it allocates the limited context budget across clusters so that different informational requirements of the query remain adequately represented. 
Third, it keeps content from the same cluster together and applies position-aware ordering across clusters within the available context. Together, these operations prevent dominant clusters from crowding out other requirements and arrange the retained information to better enable the generation model to consistently and robustly produce correct answers.

\begin{figure}[H]
    \centering
    \includegraphics[
        width=\linewidth
    ]{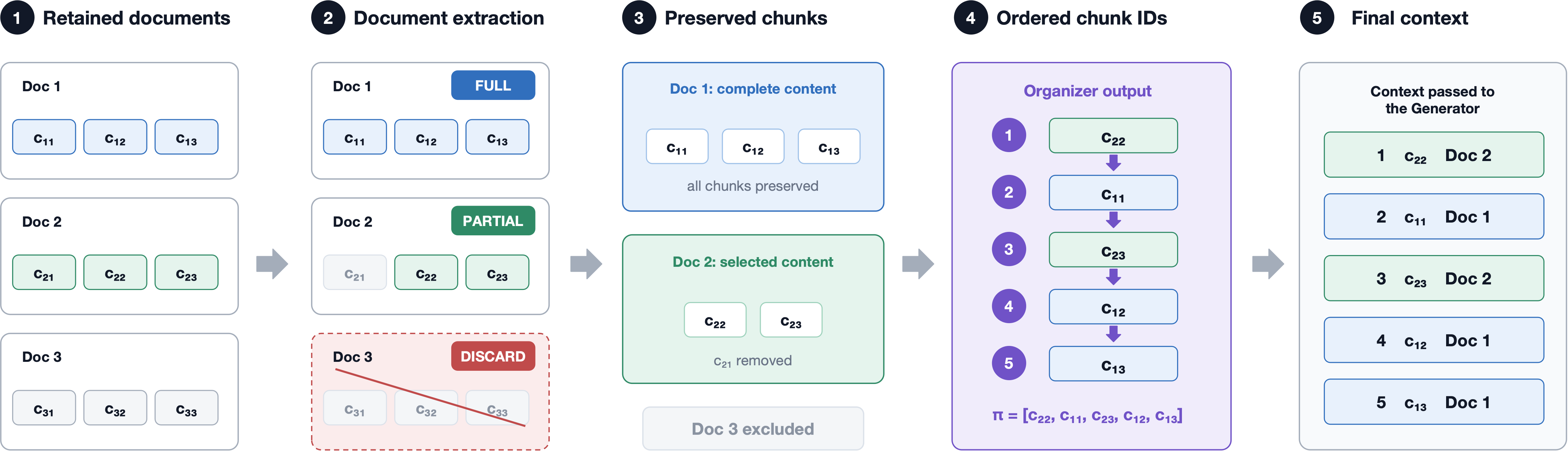}
    \caption{
    Framework of the Unified Extractor--Organizer. The model jointly performs document-level extraction and chunk-level organization to construct an ordered generation context.
    }
    \label{fig:unified-extractor-organizer}
\end{figure}

\inlinebulletsection{End-to-end extraction and organization.}
The \emph{Unified Extractor--Organizer} integrates generative information extraction with set-wise document selection and organization within a single model. Figure~\ref{fig:unified-extractor-organizer} illustrates the overall workflow, where the model first determines document-level processing decisions and then organizes the preserved content units into the final generation context.

Given a group of retained documents, it produces two levels of outputs.
At the document extraction level, the model assigns each document one of three processing decisions:
$
y_i
\in
\{
\texttt{full},
\texttt{partial},
\texttt{discard}
\}.
$
A \texttt{full} decision preserves the complete document and is appropriate when useful information is distributed throughout the document or when preserving its surrounding context is important. A \texttt{partial} decision preserves only the passages selected by the model, reducing noise and context consumption when the useful information is localized. A \texttt{discard} decision excludes a document that does not provide additional information needed for the current sub-query. 
The three-way decision allows the model to adapt the processing granularity to the informational contribution of each document rather than applying the same extraction strategy to every retained document.
At the chunk organization level, the model predicts the ordering of the preserved content units. Documents labeled \texttt{full} contribute their complete content, whereas documents labeled \texttt{partial} contribute only the selected chunks; documents labeled \texttt{discard} are excluded. The model then assigns an ordered sequence over the resulting chunk identifiers, determining how the preserved content is arranged in the final context. 

By jointly learning document-level processing decisions and chunk-level ordering, the \emph{Unified Extractor--Organizer} coordinates both what information should be preserved and how the preserved content should be organized for generation. This end-to-end formulation reduces the error propagation that can arise when generative extraction and set-wise document selection are performed by separately optimized components.

\par\medskip
\noindent\textbf{Prior-Optimization Summary.}
The complete prior-optimization workflow follows a three-stage pipeline. Candidate documents are first ranked by their estimated Answer Support scores $\widehat{\mathrm{AS}}(q,d)$, then filtered by Content Trustworthiness to form $\mathcal{C}_{\mathrm{ret}}(q)$, and finally processed jointly by Context Organization under the context budget $B$ to construct the generation context $\mathbf{X}$. In this way, prior optimization progressively transforms retrieved candidates into a bounded context whose information contributes to answer generation, is reliable for generation, and is organized to enable the generation model to consistently and robustly produce correct answers.

\subsection{Posterior Optimization in Practice}
\label{subsec:posterior-optimization-practice}

AI Search changes the granularity at which user feedback can be observed. In human-facing Web search, users interact directly with ranked documents, allowing result impressions, clicks, and post-click behavior to provide document-level supervision\cite{kim2014dwell,mehrotra2017interaction}. In AI Search, users primarily interact with the generated answer and may never access the underlying documents. Direct document-level behavioral feedback is therefore incomplete or unavailable, and the observable user outcome is concentrated at the answer level.

\begin{figure}[H]
    \centering
    \includegraphics[width=\linewidth]{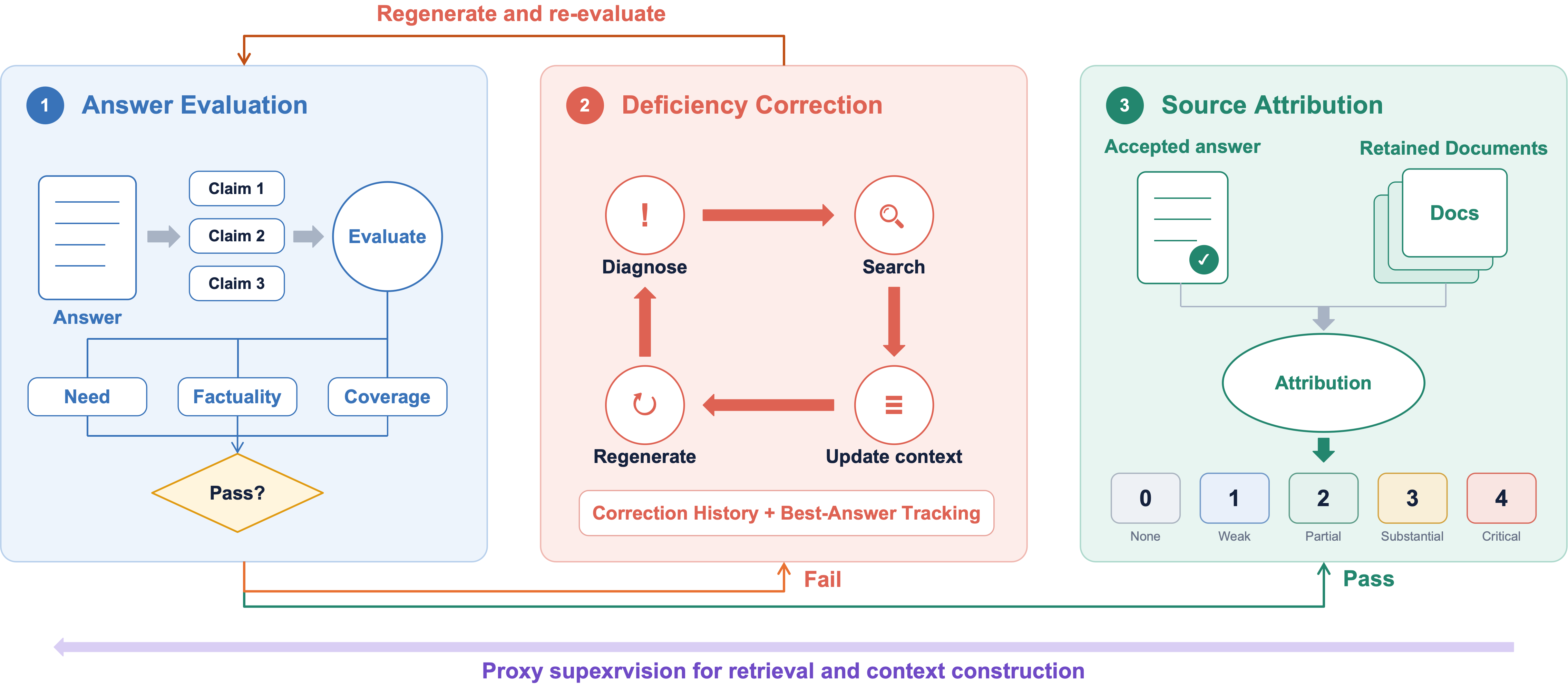}
    \caption{
    Workflow of the posterior-optimization simulator.
    The simulator evaluates generated answers, iteratively corrects diagnosed deficiencies through supplementary search, context updating, and regeneration, and attributes accepted answer outcomes to contributing documents, thereby providing proxy supervision for upstream retrieval and context-construction decisions.
    }
    \label{fig:posterior-optimization}
\end{figure}

To approximate posterior optimization under this feedback constraint, the implementation constructs a \emph{posterior-optimization simulator}. It treats the generated answer as the observable outcome of upstream retrieval and context-construction decisions and replaces currently unavailable user-feedback signals with model-based proxy judgments. The simulator operationalizes three functions: evaluating the generated answer, using diagnosed deficiencies to guide supplementary search and regeneration, and attributing successful answer outcomes back to the contributing documents~\cite{jiang2023flare,asai2024selfrag,yan2024crag}. Figure~\ref{fig:posterior-optimization} illustrates the workflow, including iterative correction for failed answers and source attribution for accepted answers.

\inlinebulletsection{Answer-level evaluation.}
Given a generated answer $\hat{y}$, the Simulator decomposes its factual content into independently verifiable claims. It generates a verification query for each claim and retrieves external evidence to classify the claim as \emph{corroborated}, \emph{contradicted}, or \emph{unresolved}. The answer is then evaluated along three dimensions:
\begin{equation}
\mathbf{s}_{\mathrm{ans}}(q,\hat{y})
=
\left(
s_{\mathrm{need}},
s_{\mathrm{fact}},
s_{\mathrm{cov}}
\right),
\end{equation}
where $s_{\mathrm{need}}$ measures whether the answer resolves the information need expressed by the query, $s_{\mathrm{fact}}$ summarizes the claim-level verification results, and $s_{\mathrm{cov}}$ measures whether the answer adequately covers the distinct perspectives or aspects required by the task. The coverage dimension is activated only for queries requiring multiple perspectives or open-ended analysis.

Evaluating the dimensions separately makes different failure modes distinguishable. Extensive analysis cannot compensate for contradicted or unresolved factual claims, and factual correctness alone does not establish that the complete information need has been addressed. The resulting score profile provides both an answer-level assessment and a structured diagnosis of uncovered requirements, factual verification failures, and missing perspectives.

\inlinebulletsection{Deficiency-guided search and regeneration.}
When an answer falls below the query-dependent acceptance threshold on any required dimension, the simulator diagnoses the remaining deficiencies, such as uncovered information requirements, contradicted or unresolved claims, and missing perspectives. Based on the original query, the current answer $\hat{y}^{(k)}$, and the history of previous correction attempts, it generates a supplementary search query targeted at these deficiencies.

Because a single supplementary search may not resolve all deficiencies, correction proceeds iteratively. Newly retrieved evidence is incorporated into the retrieval and context-construction process to reconstruct the generation context, after which the answer is regenerated and evaluated again. Each subsequent iteration focuses on the deficiencies that remain, while the correction history records previous queries, retrieved evidence, and evaluation outcomes to reduce redundant searches and repeated corrections.

The process terminates when the answer satisfies all required acceptance criteria, no further actionable deficiency can be identified, or the bounded correction budget is exhausted. Throughout the trajectory, the simulator retains the best answer according to the answer-evaluation criteria rather than assuming that the latest answer is necessarily better; the same evaluation therefore serves both as the acceptance criterion for answer selection and as a diagnostic signal for subsequent evidence acquisition and regeneration.

\inlinebulletsection{Answer-conditioned source attribution.}
For an answer that satisfies the acceptance criteria, the simulator further evaluates the contribution of each document involved in the successful generation trajectory. The attribution set includes the documents used to construct the accepted context, including additional documents introduced during correction when the accepted answer is obtained after regeneration. Acceptance is determined at the answer level and does not imply that every document in the trajectory contributes equally, or at all, to the final answer.

Each eligible document $d_i$ is assigned a five-level contribution label,
$
u_i
\in
\{0,1,2,3,4\},
$
corresponding to no, weak, partial, substantial, and critical contribution. The attribution judgment is conditioned on the accepted answer and assesses how much information from each document contributes to its generation, including whether the answer can be traced to the document, whether the document provides information not supplied by other documents in the accepted context, and whether it contributes a distinct perspective when multiple viewpoints are required. Source and temporal trustworthiness signals are retained as conditioning information during attribution, so that document contribution is assessed together with the reliability of the information it provides.

To improve attribution stability, the eligible documents are evaluated under multiple randomized source arrangements. Consistent judgments are accepted directly, while inconsistent cases are resolved through additional comparison or aggregation. The resulting label is answer-conditioned and captures each document's contribution to the accepted answer.

\smallskip
\noindent\textbf{Posterior-Optimization Summary.}
In real-world AI Search, post-generation user feedback is typically sparse, delayed, and only weakly attributable to the upstream retrieval and context-construction decisions that produced an answer. Even when explicit or implicit feedback is available, it usually reflects the overall answer outcome rather than identifying which document, trustworthiness judgment, or organization decision caused that outcome. This makes it difficult to directly construct stable supervision from observed user interactions.

We therefore implement the current Posterior-Optimization Simulator as a proxy-based workflow. Answer-level outcomes are assessed by model-based evaluators under predefined criteria, while document-contribution labels are inferred through repeated model judgments conditioned on the accepted answer and its generation trajectory. These proxy signals connect answer-level outcomes back to upstream retrieval and context-construction decisions.

\par\medskip
\noindent\textbf{Dual Optimization Perspective.}
The two paradigms provide complementary views of the same generation pipeline. Prior optimization asks whether the retrieved information contributes to answer generation, can serve as a reliable basis for generation, and is selected and organized so that the generation model can consistently and robustly produce correct answers. Posterior optimization asks whether these pre-generation decisions ultimately lead to desirable answer outcomes and uses the resulting feedback to determine which upstream decisions should be reinforced, retained, or adjusted in subsequent optimization. Together, they connect pre-generation assessment and context construction with post-generation answer outcomes, enabling iterative improvement of the overall pipeline.

\section{Evaluation Protocol and Metrics}
\label{sec:evaluation}

This section presents a reusable evaluation protocol and a corresponding set of metrics for AI Search systems. As summarized in Table~\ref{tab:evaluation-protocol}, the protocol distinguishes two evaluators---automatic and human---and two evaluation objects: the retrieval results supplied to the generation model and the generated answer delivered to the user\cite{es2024ragas,saadfalcon2024ares,ru2024ragchecker}. Automatic evaluation is applied at both the Retrieval and Answer levels, whereas human evaluation focuses on the user-facing generated answer. We first describe the resulting evaluation protocol and then formally define the metrics used in each applicable evaluator--object setting.
\begin{wraptable}[7]{r}{0.5\columnwidth}
    \centering
    \small
    \renewcommand{\arraystretch}{1.12}
    \caption{Evaluation metrics for AI Search.}
    \label{tab:evaluation-protocol}

    \begin{tabularx}{\linewidth}{
        @{}
        >{\raggedright\arraybackslash}p{0.23\linewidth}
        @{\hspace{8pt}}
        >{\raggedright\arraybackslash}X
        @{\hspace{2pt}}
        >{\raggedright\arraybackslash}p{0.23\linewidth}
        @{}
    }
        \toprule
        Evaluator & Retrieval & Answer \\
        \midrule

        Human
        & ---
        & $\mathrm{NetGain}_{\mathrm{ans}}$ \\

        Automatic
        & $\mathrm{Acc}_{qd}$,
          $\mathrm{Usable}_{q}$,
          $\mathrm{HighQ}_{q}$
        & $\mathrm{PassR}_{\mathrm{ans}}$ \\

        \bottomrule
    \end{tabularx}
\end{wraptable}

\subsection{Evaluation Protocol}
\label{subsec:evaluation-protocol}

The evaluation protocol is defined along two independent dimensions: the evaluation object and the evaluator.

\inlinebulletsection{Evaluation objects.}
We evaluate the AI Search pipeline at two checkpoints: the retrieval context supplied to answer generation and the final answer delivered to the user.

\emph{Retrieval evaluation} assesses the ranked retrieval results supplied to answer generation. Depending on the evaluation granularity, it determines either whether the result set returned for a query is usable as a whole or whether each individual query-document pair provides accurate and useful information.

\emph{Answer evaluation} assesses the generated answer presented to the user. It determines whether the final response resolves the information need after retrieval results have been selected, organized, and consumed by the generation model. Answer evaluation therefore captures the combined downstream effect of retrieval-result quality, context construction, and answer generation.

\inlinebulletsection{Evaluators.}
Automatic and human evaluation provide complementary evidence at different levels of scale and fidelity.

\emph{Automatic evaluation} uses an LLM-based evaluator to apply predefined rubrics consistently across large evaluation sets\cite{liu2023geval,zheng2023judging}. It evaluates both the retrieved information supplied to generation and the resulting answer, supporting diagnosis at the query, query--document, and answer levels. Its scalability enables rapid iteration and traffic coverage, although model-based evaluation has limited accuracy, particularly for subtle differences in answer quality.

\emph{Human evaluation} approximates the user's comparative judgment of the final output and is conducted only at the answer level. Given a query and a pair of generated answers, annotators adopt the perspective of a human user and judge which answer better addresses the query, or whether the two are comparable. Human evaluation therefore provides direct evidence of user-perceived answer quality and serves as the primary basis for assessing user-facing improvements~\cite{zheng2023judging}.

\subsection{Evaluation Metrics}
\label{subsec:evaluation-metrics}

Following the evaluator--object mapping summarized in Table~\ref{tab:evaluation-protocol}, we define the metrics used in each evaluation setting.

\subsubsection{Automatic Retrieval Evaluation}

Automatic retrieval evaluation is conducted at two granularities. Query--document-level evaluation examines the quality of each returned result independently, whereas query-level evaluation assesses whether the ranked result set collectively provides a usable or high-quality input for answer generation.

\inlinebulletsection{Query--document-level accuracy.} 
Query--document-level evaluation assesses the information supplied by each document before Context Organization, focusing on document-level properties that can be judged without considering relationships among the returned documents. 
Given query $q$ and document $d$, the evaluator first identifies the informational requirements of the query and assesses whether the document provides information that contributes to addressing those requirements. The document-level judgment then incorporates the applicable source reliability, temporal validity, and factual correctness assessments.

The evaluator assigns an ordinal label
$
a(q,d)
\in
\{0,1,2\},
$
where $0$ denotes a result that either provides no valid contribution to answer generation or fails the applicable document-level trustworthiness criteria; $1$ denotes a result that provides a valid but limited contribution to answer generation and satisfies the applicable trustworthiness criteria; and $2$ denotes a result that provides a strong contribution to answer generation and satisfies the applicable trustworthiness criteria.

The query--document-level accuracy is defined as the proportion of results that contribute to answer generation and satisfy the applicable document-level trustworthiness criteria:
\begin{equation}
\operatorname{Acc}_{qd}^{\mathrm{macro}}
=
\frac{1}{|\mathcal{Q}|}
\sum_{q\in\mathcal{Q}}
\frac{
\sum_{d\in\mathcal{D}(q)}
\mathbb{I}\!\left[a(q,d)\ge 1\right]
}{
|\mathcal{D}(q)|
},
\label{eq:acc-qu}
\end{equation}
where $\mathcal{Q}$ denotes the evaluation query set and $\mathcal{D}(q)$ denotes the documents returned for query $q$. A higher $\operatorname{Acc}_{qd}$ indicates that a larger proportion of individual results contribute to answer generation and satisfying the applicable document-level trustworthiness criteria.

\inlinebulletsection{Query-level usability and high-quality rates.}
Query-level evaluation builds on the query--document judgments above and further processes the retained documents $\mathcal{C}_{\mathrm{ret}}(q)$ to capture set-level relations that cannot be determined from individual documents alone, including cross-document contradiction identification and redundancy reduction. These operations produce revised document labels $a'(q,d)\in\{0,1,2\}$. If a document is contradicted by cross-document evidence with higher source, temporal, or factual trustworthiness, its corrected label is set to $a'(q,d)=0$. For a group of highly redundant documents, one representative retains its original label, while the remaining redundant documents are assigned $a'(q,d)=0$. All unaffected documents preserve their original labels, i.e., $a'(q,d)=a(q,d)$.


The revised document labels $a'(q,d)$ are further aggregated into a query-level score $e(q)\in\{0,1,2\}$. A query is assigned $e(q)=2$ (\emph{High-Quality Usable}) when all retained documents have labels of at least $1$ and at least one document has a label of $2$. A query is assigned $e(q)=1$ (\emph{Usable}) when at least $80\%$ of the retained documents have labels of at least $1$. Otherwise, the query is assigned $e(q)=0$ (\emph{Unusable}).
\begin{equation}
e(q)=
\begin{cases}
2, &
\displaystyle
\min_{d\in\mathcal{C}_{\mathrm{ret}}(q)} a'(q,d)\ge1
\ \land\
\max_{d\in\mathcal{C}_{\mathrm{ret}}(q)} a'(q,d)=2,
\\[6pt]
1, &
\displaystyle
\frac{
\sum_{d\in\mathcal{C}_{\mathrm{ret}}(q)}
\mathbb{I}[a'(q,d)\ge1]
}{
|\mathcal{C}_{\mathrm{ret}}(q)|
}
\ge 0.8,
\\[10pt]
0, & \text{otherwise}.
\end{cases}
\label{eq:query-score}
\end{equation}

Based on the query-level score $e(q)$, we derive the query-level usability and high-quality rates. $\operatorname{Usable}_{q}$ measures the proportion of queries that satisfy at least the \emph{Usable} criterion, i.e., $e(q)\ge1$, and $\operatorname{HighQ}_{q}$ measures the proportion that satisfy the stricter \emph{High-Quality Usable} criterion, i.e., $e(q)=2$.
\begin{equation}
\operatorname{Usable}_{q}
=
\frac{1}{|\mathcal{Q}|}
\sum_{q\in\mathcal{Q}}
\mathbb{I}[e(q)\ge1],
\qquad
\operatorname{HighQ}_{q}
=
\frac{1}{|\mathcal{Q}|}
\sum_{q\in\mathcal{Q}}
\mathbb{I}[e(q)=2].
\label{eq:query-level-metrics}
\end{equation}




\par\medskip
Taken together, query--document-level and query-level evaluation provide complementary views of retrieval performance. $\operatorname{Acc}_{qd}$ evaluates each document independently, measuring whether it contributes to answer generation and satisfies the applicable document-level trustworthiness criteria. $\operatorname{Usable}_{q}$ and $\operatorname{HighQ}_{q}$ extend the judgment to the set level by accounting for cross-document contradiction and redundancy among the retained documents. The two granularities should therefore be interpreted together.

\subsubsection{Automatic Answer Evaluation}
\label{subsubsec:automatic-answer}

Automatic Answer evaluation uses an LLM-based evaluator that takes the user query, the retrieved results, and the final generated answer as input, and assesses the answer along five dimensions:
\begin{itemize}[leftmargin=1.8em]
    \item \emph{Premise Handling} evaluates whether the answer appropriately handles erroneous, ambiguous, or underspecified premises in the query.
    \item \emph{Requirement Fulfillment} evaluates whether the answer addresses the user's stated information requirements.
    \item \emph{Factual Correctness} evaluates whether the central factual statements in the answer are consistent with the available evidence.
    \item \emph{Logical Consistency} evaluates whether the answer is internally coherent and free from contradictions.
    \item \emph{Answer Completeness} evaluates whether the answer provides sufficient task-relevant information without omitting important aspects or introducing excessive irrelevant content.
\end{itemize}

Each dimension is assigned an ordinal score in $\{0,1,2\}$, where $0$ indicates a clear or substantial failure, $1$ indicates that the criterion is basically satisfied but limitations remain, and $2$ indicates strong fulfillment of the criterion. The five dimension-level scores are aggregated by taking their minimum. An answer passes the automatic evaluation if this minimum score is at least $1$; therefore, any dimension scored $0$ causes the answer to fail.

Let $\mathcal{A}$ denote the set of evaluated answers and $\mathcal{A}_{\mathrm{pass}}$ the subset that passes the above criterion. The automatic answer pass rate is defined as
\begin{equation}
\operatorname{PassR}_{\mathrm{ans}}
=
\frac{
|\mathcal{A}_{\mathrm{pass}}|
}{
|\mathcal{A}|
}.
\label{eq:passrate}
\end{equation}
A higher $\operatorname{PassR}_{\mathrm{ans}}$ indicates that a larger proportion of generated answers satisfy the required answer criteria.

\subsubsection{Human Evaluation}
\label{subsubsec:human-evaluation}

Human evaluation assesses generated answers from the perspective of the user. Given a query and two answers produced by the evaluated system and its baseline, annotators determine which answer better satisfies the information need expressed by the query.

For each query, the two answers are presented in a blinded and randomized order and assigned a Good--Same--Bad (GSB) judgment:

\begin{itemize}[leftmargin=1.6em]
    \item \emph{Good}: the evaluated system produces the better answer;
    \item \emph{Same}: the two answers exhibit no material difference in user-perceived quality;
    \item \emph{Bad}: the evaluated system produces the worse answer.
\end{itemize}

We use GSB as the pairwise human-evaluation protocol and report Net Gain as its aggregate metric. Let $N_G$, $N_S$, and $N_B$ denote the numbers of Good, Same, and Bad judgments, respectively, with
$
N = N_G + N_S + N_B.
$
The answer-level Net Gain is defined as
\begin{equation}
\operatorname{NetGain}_{\mathrm{ans}}
=
\frac{N_G-N_B}{N}\times 100\%.
\label{eq:human-net-gain}
\end{equation}
Net Gain is the signed difference between the Good and Bad judgment rates. A positive value indicates that the system is preferred to the baseline on more queries than the baseline is preferred to the evaluated system, whereas a negative value indicates the opposite.

\section{Experiments and Case Study}
\label{sec:experiment}

In this section, we first quantitatively evaluate the implemented components associated with the three stages of the proposed framework. We then complement these results with representative case studies that illustrate how the component upgrades change system behavior and affect generated-answer quality in practice.

\begin{table*}[t]
    \centering
    \scriptsize
    \setlength{\tabcolsep}{4pt}
    \renewcommand{\arraystretch}{1.18}
    \caption{Retrieval- and answer-level effects of the implemented component upgrades. Automatic metrics are reported as absolute changes from the corresponding baselines in percentage points (pp), while human evaluation reports GSB-derived Net Gain in percent.}
    \label{tab:overall-evaluation}

    \begin{tabularx}{\textwidth}{
        l
        >{\centering\arraybackslash}X
        >{\centering\arraybackslash}X
        >{\centering\arraybackslash}X
        >{\centering\arraybackslash}X
        >{\centering\arraybackslash}X@{}
    }
        \toprule
        & \multicolumn{3}{c}{\textbf{Retrieval}}
        & \multicolumn{2}{c}{\textbf{Answer}} \\
        \cmidrule(lr){2-4}
        \cmidrule(lr){5-6}

        \textbf{Upgrade}
        & \multicolumn{3}{c}{Automatic}
        & \multicolumn{1}{c}{Automatic}
        & \multicolumn{1}{c}{Human} \\
        \cmidrule(lr){2-4}
        \cmidrule(lr){5-5}
        \cmidrule(lr){6-6}
        
        & $\mathrm{Acc}_{qu}$
        & $\mathrm{Usable}_{q}$
        & $\mathrm{HighQ}_{q}$
        & $\mathrm{PassR}_{\mathrm{ans}}$
        & $\mathrm{NetGain}_{\mathrm{ans}}$ \\
        
        \midrule
        \rowcolor{gray!12}
        \multicolumn{6}{c}{\textbf{Prior: Stage I-Answer Support}} \\
        \midrule
        
        \makecell[l]{
        Small-scale relevance ranker $\rightarrow$\\
        LLM-based Answer-Supporting ranker
        }
        & $+3.7$ pp
        & $+4.8$ pp
        & $+6.4$ pp
        & $+4.8$ pp
        & $+2.0\%$ \\
        
        \midrule
        \rowcolor{gray!12}
        \multicolumn{6}{c}{\textbf{Prior: Stage II-Content Trustworthiness}} \\
        \midrule

        Site-category $\rightarrow$ Source-role authority
        & $+2.9$ pp
        & $+3.6$ pp
        & $+4.1$ pp
        & $+5.2$ pp
        & $+2.7\%$ \\
        
        Publication-time $\rightarrow$ Content-time
        & $+2.1$ pp
        & $+1.7$ pp
        & $+3.0$ pp
        & $+4.1$ pp
        & $+3.5\%$ \\
        
        Rule-based freshness $\rightarrow$ Expiration-time
        & $+3.4$ pp
        & $+2.9$ pp
        & $+3.5$ pp
        & $+3.0$ pp
        & $+2.4\%$ \\

        N/A $\rightarrow$ Claim-level verification
        & $+2.4$ pp
        & $+2.1$ pp
        & $+4.8$ pp
        & $+3.9$ pp
        & $+4.2\%$ \\
                
        \midrule
        \rowcolor{gray!12}
        \multicolumn{6}{c}{\textbf{Prior: Stage III-Context Organization}} \\
        \midrule
        
        N/A $\rightarrow$ Extractive Extractor
        & $+5.4$ pp
        & $+4.8$ pp
        & $+6.7$ pp
        & $+4.1$ pp
        & $+8.9\%$ \\
        
        Extractive $\rightarrow$ Generative extraction
        & $+2.9$ pp
        & $+5.6$ pp
        & $+4.3$ pp
        & $+2.2$ pp
        & $+2.0\%$ \\
        
        N/A $\rightarrow$ Set-wise Organizer
        & $+4.3$ pp
        & $+7.2$ pp
        & $+4.8$ pp
        & $+1.8$ pp
        & $+1.0\%$ \\

        \midrule
        \makecell[l]{
        Extractive Extractor + Set-wise Organizer\\
        $\rightarrow$ Unified Extractor-Organizer
        }
        & $+2.2$ pp
        & $+2.1$ pp
        & $+3.6$ pp
        & $+1.5$ pp
        & $+2.8\%$ \\

        \midrule
        \rowcolor{gray!12}
        \multicolumn{6}{c}{\textbf{Posterior: Answer-Level Feedback Learning}} \\
        \midrule

        \makecell[l]{
        URL-level click supervision $\rightarrow$\\
        Sparse answer-level feedback learning
        }
        & $+1.6$ pp
        & $+2.3$ pp
        & $+3.1$ pp
        & $+4.2$ pp
        & $+4.8\%$ \\
        
        \bottomrule
    \end{tabularx}
\end{table*}

\subsection{Component-Level Evaluation}
\label{subsec:component-evaluation}

We evaluate each implemented component following the evaluation protocol and metrics defined in Section~\ref{sec:evaluation}. Each technical upgrade is compared with its corresponding baseline within the same experimental setting. For automatic evaluation, we report changes in Retrieval- and Answer-level rates in percentage points (pp). For human evaluation, we report answer-level Net Gain, expressed in percent and derived from pairwise GSB judgments. Table~\ref{tab:overall-evaluation} summarizes the component-level results. 

\subsubsection{Prior Stage I: Answer Support}
\label{subsubsec:eval-answer-supporting}

To better model the reasoning-intensive distinction between topical relevance and substantive contribution to answer generation, we replace the small-scale relevance ranker with a large-scale Answer Support ranker.
At the Retrieval level, the upgrade improves query--document accuracy by $3.7$ percentage points, query-level usability by $4.8$ percentage points, and the high-quality result-set rate by $6.4$ percentage points. These gains indicate that the upgraded ranker retains more information that contributes to subsequent answer generation. At the Answer level, the automatic pass rate increases by $4.8$ percentage points, while human evaluation reports a Net Gain of $2.0\%$.
The positive results at both evaluation levels indicate that the improvement is not confined to document-level ranking, but also translates into better generated-answer performance. Overall, the results support shifting the ranking objective beyond topical relevance toward the informational contribution that a document can make to answer generation.

\subsubsection{Prior Stage II: Content Trustworthiness}
\label{subsubsec:eval-content-trustworthiness}

\inlinebulletsection{From static source-category to domain alignment and producer profiling.}
To identify sources that are appropriate for the domain and information need of the query, we extend static source-category Authority with query-conditioned source assessment that considers Domain Alignment and Producer Profile. This upgrade raises query--document accuracy by $2.9$ percentage points, query-level usability by $3.6$ percentage points, and the high-quality result-set rate by $4.1$ percentage points. The improvement further propagates to the generated answer, increasing the automatic pass rate by $5.2$ percentage points and yielding a human-evaluated Net Gain of $2.7\%$. The gains across both evaluation levels suggest that static source-category grading alone is insufficient for model consumption: sources assigned a lower category-level Authority may still provide reliable information when they are strongly aligned with the query domain and exhibit a strong producer profile. Incorporating these signals therefore improves the reliability of the information supplied to the generation model and supports correct answer generation.

\inlinebulletsection{From publication recency to content-time assessment.}
Publication recency can be misleading when a newly published page describes an obsolete event state, expired policy, or superseded version. We therefore introduce content-time identification to determine the temporal state actually represented by the document. Under this upgrade, query--document accuracy, query-level usability, and the high-quality result-set rate improve by $2.1$, $1.7$, and $3.0$ percentage points, respectively. At the Answer level, the automatic pass rate increases by $4.1$ percentage points, accompanied by a human-evaluated Net Gain of $3.5\%$. These results show that publication time alone is an insufficient proxy for temporal applicability. By identifying the time associated with the state described in the content, the upgrade reduces the use of recently published but substantively outdated information, thereby improving the reliability of the evidence supplied for correct answer generation.

\inlinebulletsection{From rule-based freshness to expiration-time modeling.}
Fixed freshness rules cannot reflect how quickly information becomes outdated for different queries. We therefore introduce query-conditioned expiration-time modeling to estimate when the information expressed by a document should no longer be considered temporally applicable to the query. Compared with the rule-based baseline, the method improves query--document accuracy by $3.4$ percentage points and raises query-level usability and the high-quality result-set rate by $2.9$ and $3.5$ percentage points, respectively. These Retrieval-level gains translate into a $3.0$-percentage-point increase in the automatic answer pass rate and a $2.4\%$ human-evaluated Net Gain. The results indicate that temporal applicability cannot be governed by a universal recency threshold: such a threshold may retain stale information for rapidly changing queries while unnecessarily excluding valid information in more stable settings. Query-conditioned expiration modeling instead adapts the validity boundary to the temporal characteristics of the information need, improving the reliability of the evidence used for correct answer generation.

\inlinebulletsection{Claim-level verification.}
To reduce the propagation of incorrect factual claims into answers, we introduce claim-level verification that checks consequential claims against external evidence before they are used for generation. This upgrade improves query--document accuracy by $2.4$ percentage points and query-level usability by $2.1$ percentage points, while producing a larger $4.8$-percentage-point gain in the high-quality result-set rate. At the Answer level, the automatic pass rate increases by $3.9$ percentage points, and human evaluation reports a Net Gain of $4.2\%$. The stronger gains in high-quality retrieval and human-evaluated answers suggest that claim-level verification primarily improves the reliability of the retained information rather than basic answerability. By filtering or correcting factual claims that cannot be verified against external evidence, the upgrade reduces the likelihood that incorrect information propagates into the context and thereby supports correct answer generation.

\subsubsection{Prior Stage III: Context Organization}
\label{subsubsec:eval-context-organization}

\inlinebulletsection{Introducing extractive refinement.}
To remove within-document noise while preserving source-faithful, answer-supporting passages, we introduce an Extractive Extractor into a pipeline that previously had no dedicated content-refinement component. Query--document accuracy increases by $5.4$ percentage points, while query-level usability and the high-quality result-set rate increase by $4.8$ and $6.7$ percentage points, respectively. These gains propagate to a $4.1$-percentage-point increase in the automatic answer pass rate and a human-evaluated Net Gain of $8.9\%$. The improvements suggest that extractive refinement provides more than context compression: removing non-contributing content increases the density of usable evidence, while retaining passages preserves their meaning and provenance. This combination reduces the burden on the generator to locate answer-contributing evidence within noisy documents, although it remains limited when the required information cannot be represented by contiguous passages.

\inlinebulletsection{From extractive to generative extraction.}
To recover information distributed across long documents or encoded through structural relations in complex tables, we replace contiguous-passage extraction with a Generative Extractor capable of consolidating non-contiguous content. The upgrade improves query--document accuracy by $2.9$ percentage points, query-level usability by $5.6$ percentage points, and the high-quality result-set rate by $4.3$ percentage points. At the Answer level, the automatic pass rate rises by $2.2$ percentage points and human evaluation yields a Net Gain of $2.0\%$. The Retrieval-level improvements indicate that generative extraction recovers useful relations that would otherwise be fragmented or omitted by span-based processing, making previously difficult documents usable for generation. The more limited downstream gains also expose an important boundary: recovering and consolidating information does not guarantee its correct use, since the generated representation must remain faithful and must still be coordinated with evidence from other documents.

\inlinebulletsection{Introducing set-wise organization.}
To reduce cross-document redundancy while preserving complementary evidence and coverage of different information requirements, we introduce a Set-wise Organizer into a pipeline without dedicated multi-document coordination. It raises query--document accuracy by $4.3$ percentage points, query-level usability by $7.2$ percentage points, and the high-quality result-set rate by $4.8$ percentage points. The corresponding gains at the Answer level are $1.8$ percentage points in automatic pass rate and $1.0\%$ in human-evaluated Net Gain. These results indicate that document utility is not purely pointwise: information that appears redundant in isolation may provide corroboration, whereas individually useful documents may collectively overrepresent one requirement and leave others uncovered. Set-wise organization improves the balance and collective utility of the model context, but the smaller Answer-level effect shows that a well-organized evidence set remains only an enabling condition—the generator must still interpret the relationships among its content units and synthesize them correctly.

\inlinebulletsection{From separate to joint extraction and organization.}
To coordinate within-document passage extraction with cross-document selection, we replace the sequential Extractive Extractor and Set-wise Organizer with a Unified Extractor--Organizer that jointly determines which documents and passages to retain. This joint formulation improves query--document accuracy, query-level usability, and the high-quality result-set rate by $2.2$, $2.1$, and $3.6$ percentage points, respectively. It also increases the automatic answer pass rate by $1.5$ percentage points and produces a human-evaluated Net Gain of $2.8\%$. In a sequential pipeline, the extractor may remove content that appears weak at the document level but becomes complementary or necessary when considered alongside other candidates; the downstream organizer cannot recover information that has already been discarded. Joint modeling allows passage-level retention to account for set-level contribution, thereby reducing this form of irreversible error propagation. The positive Retrieval- and Answer-level results support coordinating the two decisions under a shared representation rather than optimizing them independently.

\subsubsection{Posterior Optimization: Answer-Level Feedback Learning}
\label{subsubsec:eval-posterior-optimization}

To better align optimization with the final answer outcome, we replace document-level click supervision with sparse answer-level feedback learning. This upgrade improves query--document accuracy by $1.6$ percentage points, query-level usability by $2.3$ percentage points, and the high-quality result-set rate by $3.1$ percentage points. The gains are more pronounced at the Answer level, where the automatic pass rate increases by $4.2$ percentage points and human evaluation reports a Net Gain of $4.8\%$.

The greater improvements at the Answer level are consistent with the objective of posterior optimization: rather than relying only on document-level behavioral signals, answer-level feedback directly reflects whether the final response succeeds and can be attributed back to the documents and context-construction decisions that contributed to that outcome. This enables the system to learn from sparse but more outcome-aligned supervision, improving the upstream retrieval and context decisions in ways that are ultimately reflected in better generated answers.

\subsection{Case Study}
\label{subsec:case-study}

To complement the quantitative evaluation, we present representative case studies that illustrate how the three-stage framework improves the generation context in practice. The cases show how Answer Support identifies information needed for answer generation, Content Trustworthiness filters unreliable evidence to enable correct answer generation, and Context Organization restructures retained information for more consistent and robust generation. We further analyze a failure case that exposes a limitation of the current implementation and motivates future improvement.

\inlinebulletsection{Case 1: Identifying Implicit Derivational Evidence.}
Given the query, ``What is the latest time to register an ICLR 2027 submission?'', the small-scale relevance ranker prioritizes \emph{ICLR 2027 Dates and Deadlines} because it directly matches the explicit conference and submission-time terms. This document establishes that the abstract deadline is September 18, 2026, at 23:59 AoE.

The LLM-based Answer-Supporting ranker reasons beyond the terms explicitly stated in the query. It recognizes that resolving the requested deadline requires determining the user's local time zone and converting the AoE deadline accordingly. It incorporates the user's location, Beijing, and further retrieves documents describing the Beijing time zone and its conversion from AoE. By including additional documents containing the information required to derive the answer, the LLM-based ranker provides a more complete basis for generation and thereby helps improve the quality of the final answer.

\begin{wrapfigure}[16]{r}{0.48\columnwidth}
    \centering
    \includegraphics[
        width=\linewidth
    ]{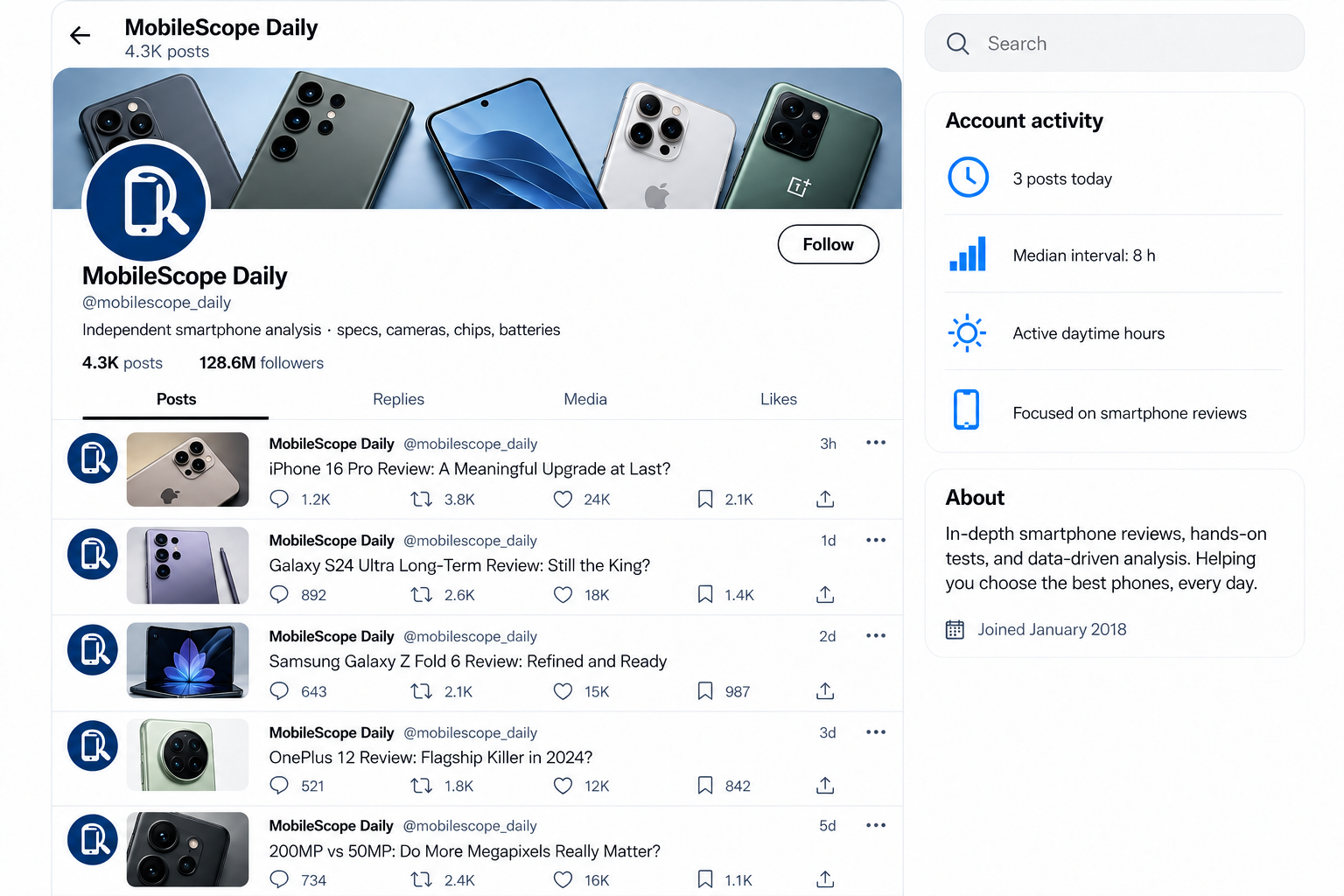}
    \caption{
    Strong query--producer domain alignment and a healthy producer profile can identify reliable sources that may be underestimated by static source-category Authority.
    }
    \label{fig:producer}
\end{wrapfigure}

\inlinebulletsection{Case 2: Recognizing High-Quality Domain-Specialized Producers.}
Figure~\ref{fig:producer} considers the query, ``Is the iPhone 16 Pro worth buying?'' Under the conventional Authority formulation, an individual publisher receives a relatively low authority level because the assessment is determined primarily by the predefined source category. However, the producer in this example is a well-established smartphone reviewer with more than 100 million followers and a long publication history focused on mobile devices. 
The current Source Trustworthiness assessment further evaluates both the alignment between the query domain and the producer's area of specialization, and the quality of the producer profile. In this case, the query concerns a smartphone purchase decision, while the producer has repeatedly reviewed smartphones and successive generations of the iPhone, resulting in strong domain alignment. At the same time, the producer's large audience, sustained publication history, consistently strong engagement, and regular publishing behavior indicate a stable and high-quality producer profile. Together, these signals indicate that the creator can serve as an appropriate and reliable source for the query. Incorporating such sources provides the generation model with higher-quality evidence and helps support correct answer generation.

\inlinebulletsection{Case 3: Identifying Temporally Applicable Content Beyond Publication Recency.}
At the search time of August 30, 2026, the query asks, ``Who is Luckin Coffee's spokesperson?'' DocA, published on August 14, 2026, states that \emph{Luckin Coffee officially signed freestyle-skiing world champion Eileen Gu as a brand ambassador}. Under the conventional Freshness signal, DocA receives a strong freshness weight because it was published within the preceding three months. However, Temporal Trustworthiness further examines when the reported information actually applies and whether it remains valid at the current search time. Through real-time retrieval and temporal verification, the system identifies the content time of DocA as 2021 and confirms that the endorsement of Eileen Gu described in the document is no longer current.
DocB, entitled \emph{Official Announcement: Liu Yifei Becomes Luckin Coffee's Global Brand Ambassador and Chief Recommender for Tea Beverages}, was published on August 12, 2024. Under the publication-time-based Freshness criterion, it therefore receives a lower freshness assessment. Temporal Trustworthiness, however, verifies that the information reported in DocB remains applicable at the current search time, confirming that Liu Yifei remains a Luckin Coffee brand ambassador. DocB is therefore promoted as the temporally valid evidence despite its earlier publication date.
By decoupling temporal validity from publication recency, Temporal Trustworthiness improves the reliability of the evidence supplied to the generation model, helping prevent outdated information from propagating into incorrect answers.

\inlinebulletsection{Case 4: Organizing Complementary Evidence into a Structured Context.}
Consider the query, ``I successively contributed to pension insurance in Beijing, Shanghai, and Shangrao, Jiangxi, for fewer than ten years in each location. Where should I claim retirement benefits, and how will my contribution records be calculated?'' The four official documents have passed the preceding Answer-Supporting and Content Trustworthiness assessments. Doc~1 establishes the national rules for determining the benefit-claim location and aggregating contribution records; Doc~2 clarifies the consolidation responsibility of the household-registration location and several exceptional situations; Doc~3 provides an operational explanation of cross-province relationship, fund, and record transfer; and Doc~4 clarifies how contribution records are handled after transfer, including the accumulation of non-overlapping contribution periods and the treatment of duplicate contributions.

Directly concatenating the four documents exposes the Generator to both repeated provisions and details that do not affect the stated case. Doc~1 includes fund-transfer formulas, processing deadlines, and information-system requirements; Doc~2 discusses missing historical records, lump-sum contributions, duplicate benefit receipt, veterans, and enterprise relocation; Doc~3 includes age-conditioned temporary-account rules and detailed fund-transfer calculations; and Doc~4 contains additional transfer-related details beyond the contribution-record rules needed for the query. Although valid within their respective scopes, these passages are unnecessary for determining the benefit-claim location or aggregating the user's contribution records. Retaining them consumes context budget, obscures key decision conditions, and may introduce inapplicable exceptions or confuse record aggregation with fund-transfer calculation.

In the Context Organization workflow, the Extractor first removes content within individual documents that does not contribute to the requested answer. The Organizer then removes repeated provisions across documents and groups complementary information from different documents by the answer aspect it addresses. For example, the benefit-location rules from Doc~1 and Doc~2 are consolidated into one evidence group, while the contribution-record rules from Doc~1 and Doc~4 are organized together into another. The resulting evidence groups are then ordered according to the answer logic, from the applicable condition to the benefit-claim location and finally to contribution-record aggregation. This organization transforms a document-wise sequence into a topic-structured context in which complementary information from different sources is colocated for model consumption.

The resulting context enables the Generator to use the retained evidence more consistently and robustly. In this case, the benefits should be claimed at the household-registration location, while eligible records from Beijing, Shanghai, and Jiangxi are accumulated. Because the query does not specify that location, the answer should state the governing rule rather than incorrectly selecting one of the three contribution locations. By reducing irrelevant content, eliminating redundancy, and colocating complementary information, Context Organization supports the consistent and robust generation of the correct answer.

\begin{wrapfigure}[14]{r}{0.45\columnwidth}
    \centering
    \includegraphics[
        width=\linewidth
    ]{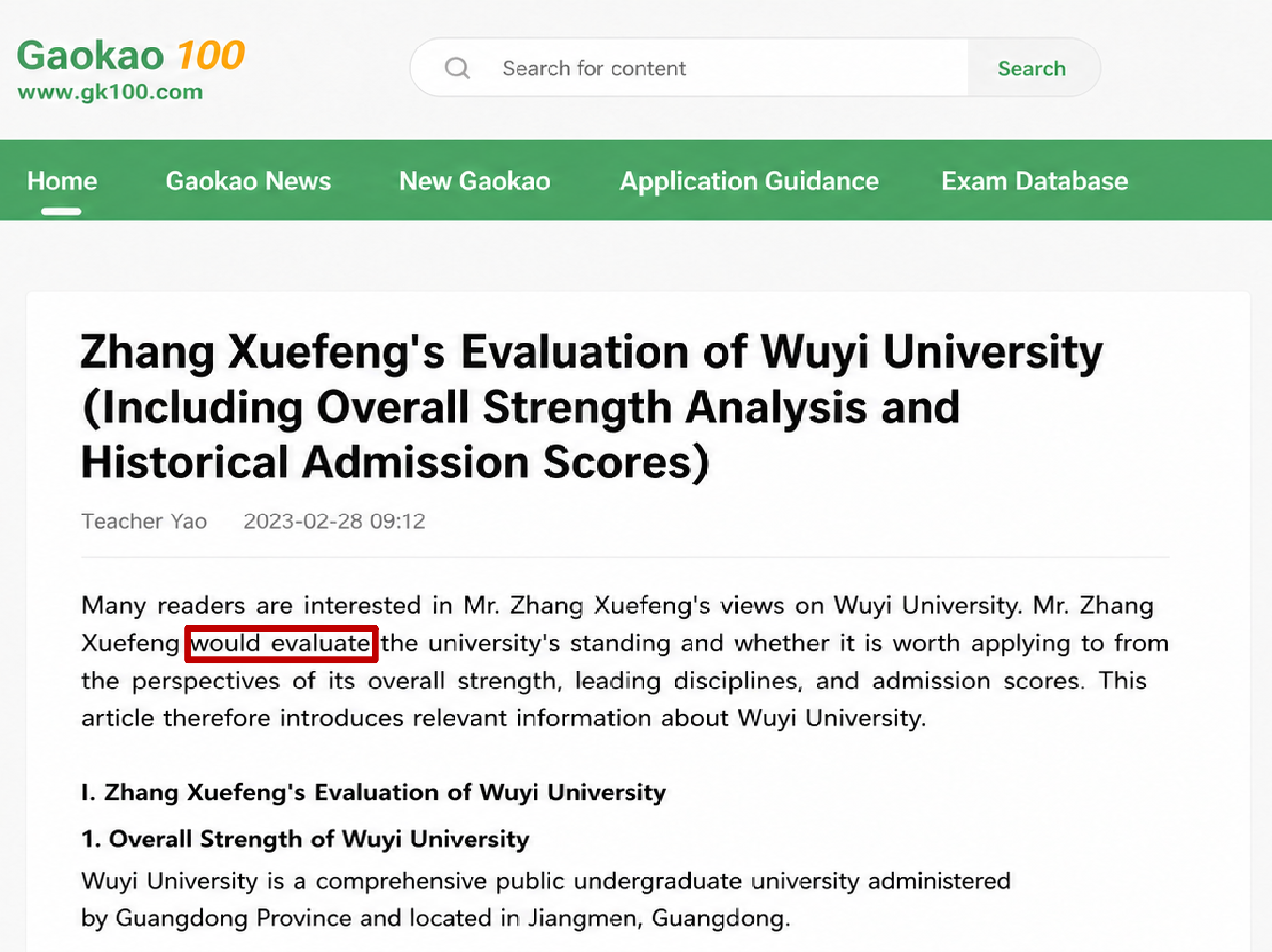}
    \caption{A failure case for claim-level verification involving unsupported attribution.}
    \label{fig:claim-verification-badcase}
\end{wrapfigure}
\inlinebulletsection{Case 5: Failure to Detect Mutually Reinforced False Claims.}
Figure~\ref{fig:claim-verification-badcase} presents one example of a broader failure mode for the query, ``How does Zhang Xuefeng evaluate Wuyi University?'' The Web contains multiple documents that attribute similar evaluations of Wuyi University to Zhang Xuefeng, and the example shown here is one such document. However, the attributed claim is not grounded in a verifiable statement made by him and is therefore false.
Because similar claims are repeated across multiple documents, the retrieved evidence may appear to corroborate itself. These documents may appear to corroborate one another, even though they merely repeat the same false claim rather than provide independent evidence for it. The current Information Trustworthiness implementation may fail to distinguish this repeated agreement from genuine verification and therefore retain the claim as factually trustworthy.
The Generator may consequently reproduce the false attribution as Zhang Xuefeng's actual view. This failure shows that Claim-Level Verification must determine whether multiple documents provide genuinely independent evidence for a claim, rather than simply repeating the same false information, which remains an important direction for future work.

\section{Conclusion}

This paper revisits the role of retrieval in AI Search from the perspective of a fundamental shift in information consumption: retrieved documents are no longer end products presented directly to users, but instead serve as inputs to the generation model. Under this shift, the human-facing Search Satisfaction objective is insufficient to characterize the requirements of constructing context for answer generation. 
We therefore establish a three-stage view of retrieval in AI Search through \emph{Answer Support}, \emph{Content Trustworthiness}, and \emph{Context Organization}. Together, these stages progressively examine whether retrieved content contributes to answer generation, whether it provides a reliable basis for correct answers, and how the retained information should be organized into a bounded context to enable correct generation consistently and robustly

We further operationalize this framework through an industrial workflow comprising prior and posterior optimization. Prior optimization implements the three stages before generation, while posterior optimization uses answer-level outcomes to provide feedback on preceding retrieval and context-construction decisions. To evaluate this process systematically, we establish a protocol covering both the retrieval results supplied to the generation model and the generated answers delivered to users, combining scalable automatic evaluation with human assessment of user-facing answer quality. Online experiments show that the implemented component upgrades consistently improve both Retrieval- and Answer-level performance, demonstrating that improving the information supplied to the model can translate into better final answers. Taken together, our findings highlight the need to move beyond ranking satisfactory documents and treat post-retrieval processing as the construction of reliable, well-organized context for answer generation.

\bibliographystyle{IEEEtran}
\bibliography{reference}

@inproceedings{lewis2020rag,
  title = {Retrieval-Augmented Generation for Knowledge-Intensive {NLP} Tasks},
  author = {Lewis, Patrick and Perez, Ethan and Piktus, Aleksandra and Petroni, Fabio and Karpukhin, Vladimir and Goyal, Naman and K{\"u}ttler, Heinrich and Lewis, Mike and Yih, Wen-tau and Rockt{\"a}schel, Tim and Riedel, Sebastian and Kiela, Douwe},
  booktitle = {Advances in Neural Information Processing Systems},
  volume = {33},
  pages = {9459--9474},
  year = {2020},
  url = {https://proceedings.neurips.cc/paper/2020/hash/6b493230205f780e1bc26945df7481e5-Abstract.html},
  eprint = {2005.11401},
  archiveprefix = {arXiv},
  primaryclass = {cs.CL},
}

@inproceedings{zhang2026beyond,
  title = {Beyond Relevance: Utility-Centric Retrieval in the LLM Era},
  author = {Zhang, Hengran and Tang, Minghao and Bi, Keping and Guo, Jiafeng},
  booktitle = {Proceedings of the 49th International ACM SIGIR Conference on Research and Development in Information Retrieval},
  pages = {5357--5361},
  year = {2026},
  publisher = {ACM},
  doi = {10.1145/3805712.3808641},
  url = {https://doi.org/10.1145/3805712.3808641},
}

@inproceedings{dai2025seper,
  title = {{SePer}: Measure Retrieval Utility Through the Lens of Semantic Perplexity Reduction},
  author = {Dai, Lu and Xu, Yijie and Ye, Jinhui and Liu, Hao and Xiong, Hui},
  booktitle = {The Thirteenth International Conference on Learning Representations},
  year = {2025},
  url = {https://openreview.net/forum?id=ixMBnOhFGd},
}

@inproceedings{lee2025setselection,
  title = {Shifting from Ranking to Set Selection for Retrieval Augmented Generation},
  author = {Lee, Dahyun and Jo, Yongrae and Park, Haeju and Lee, Moontae},
  booktitle = {Proceedings of the 63rd Annual Meeting of the Association for Computational Linguistics (Volume 1: Long Papers)},
  pages = {17606--17619},
  year = {2025},
  publisher = {Association for Computational Linguistics},
  doi = {10.18653/v1/2025.acl-long.861},
  url = {https://aclanthology.org/2025.acl-long.861/},
}

@article{liu2024lost,
  title = {Lost in the Middle: How Language Models Use Long Contexts},
  author = {Liu, Nelson F. and Lin, Kevin and Hewitt, John and Paranjape, Ashwin and Bevilacqua, Michele and Petroni, Fabio and Liang, Percy},
  journal = {Transactions of the Association for Computational Linguistics},
  volume = {12},
  pages = {157--173},
  year = {2024},
  doi = {10.1162/tacl_a_00638},
  url = {https://aclanthology.org/2024.tacl-1.9/},
}

@inproceedings{jiang2024longllmlingua,
  title = {{LongLLMLingua}: Accelerating and Enhancing {LLM}s in Long Context Scenarios via Prompt Compression},
  author = {Jiang, Huiqiang and Wu, Qianhui and Luo, Xufang and Li, Dongsheng and Lin, Chin-Yew and Yang, Yuqing and Qiu, Lili},
  booktitle = {Proceedings of the 62nd Annual Meeting of the Association for Computational Linguistics (Volume 1: Long Papers)},
  pages = {1658--1677},
  year = {2024},
  address = {Bangkok, Thailand},
  publisher = {Association for Computational Linguistics},
  doi = {10.18653/v1/2024.acl-long.91},
  url = {https://aclanthology.org/2024.acl-long.91/},
}

@inproceedings{xu2023recomp,
  title = {{RECOMP}: Improving Retrieval-Augmented {LM}s with Compression and Selective Augmentation},
  author = {Xu, Fangyuan and Shi, Weijia and Choi, Eunsol},
  booktitle = {The Twelfth International Conference on Learning Representations},
  year = {2024},
  url = {https://openreview.net/forum?id=mlJLVigNHp},
}

@inproceedings{yoran2023irrelevant,
  title = {Making Retrieval-Augmented Language Models Robust to Irrelevant Context},
  author = {Yoran, Ori and Wolfson, Tomer and Ram, Ori and Berant, Jonathan},
  booktitle = {The Twelfth International Conference on Learning Representations},
  year = {2024},
  url = {https://openreview.net/forum?id=ZS4m74kZpH},
}

@inproceedings{joren2024sufficient,
  title = {Sufficient Context: A New Lens on Retrieval Augmented Generation Systems},
  author = {Joren, Hailey and Zhang, Jianyi and Ferng, Chun-Sung and Juan, Da-Cheng and Taly, Ankur and Rashtchian, Cyrus},
  booktitle = {The Thirteenth International Conference on Learning Representations},
  year = {2025},
  url = {https://openreview.net/forum?id=Jjr2Odj8DJ},
}

@inproceedings{hwang2024rarag,
  title = {Retrieval-Augmented Generation with Estimation of Source Reliability},
  author = {Hwang, Jeongyeon and Park, Junyoung and Park, Hyejin and Kim, Dongwoo and Park, Sangdon and Ok, Jungseul},
  booktitle = {Proceedings of the 2025 Conference on Empirical Methods in Natural Language Processing},
  pages = {34279--34303},
  year = {2025},
  publisher = {Association for Computational Linguistics},
  doi = {10.18653/v1/2025.emnlp-main.1738},
  url = {https://aclanthology.org/2025.emnlp-main.1738/},
}

@misc{jin2026sourcebench,
  title = {{SourceBench}: Can {AI} Answers Reference Quality Web Sources?},
  author = {Jin, Hexi and Liu, Stephen and Li, Yuheng and Malik, Simran and Zhang, Yiying},
  year = {2026},
  url = {https://arxiv.org/abs/2602.16942},
  eprint = {2602.16942},
  archiveprefix = {arXiv},
  primaryclass = {cs.CL},
}

@inproceedings{gao2023alce,
  title = {Enabling Large Language Models to Generate Text with Citations},
  author = {Gao, Tianyu and Yen, Howard and Yu, Jiatong and Chen, Danqi},
  booktitle = {Proceedings of the 2023 Conference on Empirical Methods in Natural Language Processing},
  pages = {6465--6488},
  year = {2023},
  address = {Singapore},
  publisher = {Association for Computational Linguistics},
  doi = {10.18653/v1/2023.emnlp-main.398},
  url = {https://aclanthology.org/2023.emnlp-main.398/},
}

@inproceedings{niu2024ragtruth,
  title = {{RAGT}ruth: A Hallucination Corpus for Developing Trustworthy Retrieval-Augmented Language Models},
  author = {Niu, Cheng and Wu, Yuanhao and Zhu, Juno and Xu, Siliang and Shum, KaShun and Zhong, Randy and Song, Juntong and Zhang, Tong},
  booktitle = {Proceedings of the 62nd Annual Meeting of the Association for Computational Linguistics (Volume 1: Long Papers)},
  pages = {10862--10878},
  year = {2024},
  address = {Bangkok, Thailand},
  publisher = {Association for Computational Linguistics},
  doi = {10.18653/v1/2024.acl-long.585},
  url = {https://aclanthology.org/2024.acl-long.585/},
}

@misc{dervin1983sensemaking,
  title = {An Overview of Sense-Making Research: Concepts, Methods, and Results to Date},
  author = {Dervin, Brenda},
  year = {1983},
  url = {https://books.google.com/books?id=wInhAAAAMAAJ},
  note = {Presented at the International Communication Association Annual Meeting},
}

@article{wilson1981userstudies,
  title = {On User Studies and Information Needs},
  author = {Wilson, T. D.},
  journal = {Journal of Documentation},
  volume = {37},
  number = {1},
  pages = {3--15},
  year = {1981},
  doi = {10.1108/eb026702},
  url = {https://doi.org/10.1108/eb026702},
}

@article{brin1998anatomy,
  title = {The Anatomy of a Large-Scale Hypertextual Web Search Engine},
  author = {Brin, Sergey and Page, Lawrence},
  journal = {Computer Networks and ISDN Systems},
  volume = {30},
  number = {1--7},
  pages = {107--117},
  year = {1998},
  doi = {10.1016/S0169-7552(98)00110-X},
  url = {https://doi.org/10.1016/S0169-7552(98)00110-X},
}

@article{yang2005webir,
  title = {Information Retrieval on the Web},
  author = {Yang, Kiduk},
  journal = {Annual Review of Information Science and Technology},
  volume = {39},
  number = {1},
  pages = {33--80},
  year = {2005},
  doi = {10.1002/aris.1440390109},
}

@inproceedings{thorne2018fever,
  title = {{FEVER}: A Large-Scale Dataset for Fact Extraction and {VER}ification},
  author = {Thorne, James and Vlachos, Andreas and Christodoulopoulos, Christos and Mittal, Arpit},
  booktitle = {Proceedings of the 2018 Conference of the North American Chapter of the Association for Computational Linguistics: Human Language Technologies, Volume 1 (Long Papers)},
  pages = {809--819},
  year = {2018},
  publisher = {Association for Computational Linguistics},
  doi = {10.18653/v1/N18-1074},
  url = {https://aclanthology.org/N18-1074/},
}

@inproceedings{min2023factscore,
  title = {{FActScore}: Fine-Grained Atomic Evaluation of Factual Precision in Long-Form Text Generation},
  author = {Min, Sewon and Krishna, Kalpesh and Lyu, Xinxi and Lewis, Mike and Yih, Wen-tau and Koh, Pang and Iyyer, Mohit and Zettlemoyer, Luke and Hajishirzi, Hannaneh},
  booktitle = {Proceedings of the 2023 Conference on Empirical Methods in Natural Language Processing},
  pages = {12076--12100},
  year = {2023},
  publisher = {Association for Computational Linguistics},
  doi = {10.18653/v1/2023.emnlp-main.741},
  url = {https://aclanthology.org/2023.emnlp-main.741/},
}

@inproceedings{guo2017deepfm,
  title     = {DeepFM: A Factorization-Machine Based Neural Network for CTR Prediction},
  author    = {Guo, Huifeng and Tang, Ruiming and Ye, Yunming and Li, Zhenguo and He, Xiuqiang},
  booktitle = {Proceedings of the Twenty-Sixth International Joint Conference on Artificial Intelligence},
  pages     = {1725--1731},
  year      = {2017},
  doi       = {10.24963/ijcai.2017/239}
}

@inproceedings{devlin2019bert,
  title = {{BERT}: Pre-training of Deep Bidirectional Transformers for Language Understanding},
  author = {Devlin, Jacob and Chang, Ming-Wei and Lee, Kenton and Toutanova, Kristina},
  booktitle = {Proceedings of the 2019 Conference of the North American Chapter of the Association for Computational Linguistics: Human Language Technologies, Volume 1 (Long and Short Papers)},
  pages = {4171--4186},
  year = {2019},
  publisher = {Association for Computational Linguistics},
  doi = {10.18653/v1/N19-1423},
  url = {https://aclanthology.org/N19-1423/},
}

@inproceedings{
sun2026rethinking,
title={Rethinking the Reranker: Boundary-Aware Evidence Selection for Robust Retrieval-Augmented Generation},
author={Jiashuo Sun and Pengcheng Jiang and Saizhuo Wang and Jiajun Fan and Heng Wang and Siru Ouyang and Ming Zhong and Yizhu Jiao and Chengsong Huang and Xueqiang Xu and Pengrui Han and Peiran Li and Jiaxin Huang and Ge Liu and Heng Ji and Jiawei Han},
booktitle={Forty-third International Conference on Machine Learning},
year={2026},
url={https://openreview.net/forum?id=Tt8lCe1NrW}
}

@misc{ai2023informationretrievalmeetslarge,
      title={Information Retrieval Meets Large Language Models: A Strategic Report from Chinese IR Community}, 
      author={Qingyao Ai and Ting Bai and Zhao Cao and Yi Chang and Jiawei Chen and Zhumin Chen and Zhiyong Cheng and Shoubin Dong and Zhicheng Dou and Fuli Feng and Shen Gao and Jiafeng Guo and Xiangnan He and Yanyan Lan and Chenliang Li and Yiqun Liu and Ziyu Lyu and Weizhi Ma and Jun Ma and Zhaochun Ren and Pengjie Ren and Zhiqiang Wang and Mingwen Wang and Ji-Rong Wen and Le Wu and Xin Xin and Jun Xu and Dawei Yin and Peng Zhang and Fan Zhang and Weinan Zhang and Min Zhang and Xiaofei Zhu},
      year={2023},
      eprint={2307.09751},
      archivePrefix={arXiv},
      primaryClass={cs.IR},
      url={https://arxiv.org/abs/2307.09751}, 
}

@article{hilligoss2008developing,
  title   = {Developing a Unifying Framework of Credibility Assessment: Construct, Heuristics, and Interaction in Context},
  author  = {Hilligoss, Brian and Rieh, Soo Young},
  journal = {Information Processing \& Management},
  volume  = {44},
  number  = {4},
  pages   = {1467--1484},
  year    = {2008},
  doi     = {10.1016/j.ipm.2007.10.001}
}

@inproceedings{trappolini-etal-2026-redefining,
    title = "Redefining Retrieval Evaluation in the Era of {LLM}s",
    author = "Trappolini, Giovanni  and
      Cuconasu, Florin  and
      Filice, Simone  and
      Maarek, Yoelle  and
      Silvestri, Fabrizio",
    editor = "Demberg, Vera  and
      Inui, Kentaro  and
      Marquez, Llu{\'i}s",
    booktitle = "Proceedings of the 19th Conference of the {E}uropean Chapter of the {A}ssociation for {C}omputational {L}inguistics (Volume 1: Long Papers)",
    month = mar,
    year = "2026",
    address = "Rabat, Morocco",
    publisher = "Association for Computational Linguistics",
    url = "https://aclanthology.org/2026.eacl-long.391/",
    doi = "10.18653/v1/2026.eacl-long.391",
    pages = "8359--8375",
    ISBN = "979-8-89176-380-7"
}

@inproceedings{fan2024ragsurvey,
  title = {A Survey on {RAG} Meeting {LLM}s: Towards Retrieval-Augmented Large Language Models},
  author = {Fan, Wenqi and Ding, Yujuan and Ning, Liangbo and Wang, Shijie and Li, Hengyun and Yin, Dawei and Chua, Tat-Seng and Li, Qing},
  booktitle = {Proceedings of the 30th ACM SIGKDD Conference on Knowledge Discovery and Data Mining},
  pages = {6491--6501},
  year = {2024},
  publisher = {Association for Computing Machinery},
  doi = {10.1145/3637528.3671470},
  url = {https://doi.org/10.1145/3637528.3671470},
}

@inproceedings{salemi2024searchmachines,
  title = {Towards a Search Engine for Machines: Unified Ranking for Multiple Retrieval-Augmented Large Language Models},
  author = {Salemi, Alireza and Zamani, Hamed},
  booktitle = {Proceedings of the 47th International ACM SIGIR Conference on Research and Development in Information Retrieval},
  pages = {741--751},
  year = {2024},
  doi = {10.1145/3626772.3657733},
  url = {https://doi.org/10.1145/3626772.3657733},
}

@inproceedings{asai2024selfrag,
  title = {{Self-RAG}: Learning to Retrieve, Generate, and Critique through Self-Reflection},
  author = {Asai, Akari and Wu, Zeqiu and Wang, Yizhong and Sil, Avirup and Hajishirzi, Hannaneh},
  booktitle = {The Twelfth International Conference on Learning Representations},
  year = {2024},
  url = {https://openreview.net/forum?id=hSyW5go0v8},
}

@inproceedings{morlan2026factappeal,
  title = {{FactAppeal}: Identifying Epistemic Factual Appeals in News Media},
  author = {Mor-Lan, Guy and Sheafer, Tamir and Shenhav, Shaul R.},
  booktitle = {Findings of the Association for Computational Linguistics: EACL 2026},
  pages = {6545--6556},
  year = {2026},
  publisher = {Association for Computational Linguistics},
  doi = {10.18653/v1/2026.findings-eacl.344},
  url = {https://aclanthology.org/2026.findings-eacl.344/},
}

@inproceedings{zhang2021situatedqa,
  title = {{SituatedQA}: Incorporating Extra-Linguistic Contexts into {QA}},
  author = {Zhang, Michael and Choi, Eunsol},
  booktitle = {Proceedings of the 2021 Conference on Empirical Methods in Natural Language Processing},
  pages = {7371--7387},
  year = {2021},
  publisher = {Association for Computational Linguistics},
  doi = {10.18653/v1/2021.emnlp-main.586},
  url = {https://aclanthology.org/2021.emnlp-main.586/},
}

@inproceedings{cao2026re3,
  title = {Re$^3$: Relevance {\&} Recency Retrieval for Mitigating Temporal Hallucination},
  author = {Cao, Jiawei and Ouyang, Jie and Cheng, Mingyue and Zhou, Zhaomeng and Liu, Chunli and Li, Yupeng and Liu, Zirui and Wang, Shijin},
  booktitle = {Proceedings of the 64th Annual Meeting of the Association for Computational Linguistics},
  pages = {25735--25760},
  year = {2026},
  publisher = {Association for Computational Linguistics},
  doi = {10.18653/v1/2026.acl-long.1180},
  url = {https://aclanthology.org/2026.acl-long.1180/},
}

@inproceedings{ouyang2025hoh,
  title = {{HoH}: A Dynamic Benchmark for Evaluating the Impact of Outdated Information on Retrieval-Augmented Generation},
  author = {Ouyang, Jie and Pan, Tingyue and Cheng, Mingyue and Yan, Ruiran and Luo, Yucong and Lin, Jiaying and Liu, Qi},
  booktitle = {Proceedings of the 63rd Annual Meeting of the Association for Computational Linguistics},
  pages = {6036--6063},
  year = {2025},
  publisher = {Association for Computational Linguistics},
  doi = {10.18653/v1/2025.acl-long.301},
  url = {https://aclanthology.org/2025.acl-long.301/},
}

@inproceedings{schlichtkrull2023averitec,
  title = {{AVeriTeC}: A Dataset for Real-world Claim Verification with Evidence from the Web},
  author = {Schlichtkrull, Michael and Guo, Zhijiang and Vlachos, Andreas},
  booktitle = {Advances in Neural Information Processing Systems},
  volume = {36},
  year = {2023},
  url = {https://proceedings.neurips.cc/paper_files/paper/2023/hash/cd86a30526cd1aff61d6f89f107634e4-Abstract-Datasets_and_Benchmarks.html},
}

@inproceedings{chen2026expiration,
  title = {{RAG}-Enhanced Large Language Models for Dynamic Content Expiration Prediction in Web Search},
  author = {Chen, Tingyu and Zhang, Wenkai and Gao, Li and Su, Lixin and Chen, Ge and Yin, Dawei and Shi, Daiting},
  booktitle = {Proceedings of the 49th International ACM SIGIR Conference on Research and Development in Information Retrieval},
  pages = {4523--4527},
  year = {2026},
  publisher = {ACM},
  doi = {10.1145/3805712.3808457},
  url = {https://doi.org/10.1145/3805712.3808457},
}

@inproceedings{mao2016relevance,
  title = {When Does Relevance Mean Usefulness and User Satisfaction in Web Search?},
  author = {Mao, Jiaxin and Liu, Yiqun and Zhou, Ke and Nie, Jian-Yun and Song, Jingtao and Zhang, Min and Ma, Shaoping and Sun, Jiashen and Luo, Hengliang},
  booktitle = {Proceedings of the 39th International ACM SIGIR Conference on Research and Development in Information Retrieval},
  pages = {463--472},
  year = {2016},
  publisher = {ACM},
  doi = {10.1145/2911451.2911507},
}

@inproceedings{kim2014dwell,
  title = {Modeling Dwell Time to Predict Click-Level Satisfaction},
  author = {Kim, Youngho and Hassan Awadallah, Ahmed and White, Ryen W. and Zitouni, Imed},
  booktitle = {Proceedings of the 7th ACM International Conference on Web Search and Data Mining},
  pages = {193--202},
  year = {2014},
  publisher = {ACM},
  doi = {10.1145/2556195.2556220},
}

@inproceedings{mehrotra2017interaction,
  title = {User Interaction Sequences for Search Satisfaction Prediction},
  author = {Mehrotra, Rishabh and Zitouni, Imed and Hassan Awadallah, Ahmed and El Kholy, Ahmed and Khabsa, Madian},
  booktitle = {Proceedings of the 40th International ACM SIGIR Conference on Research and Development in Information Retrieval},
  pages = {165--174},
  year = {2017},
  publisher = {ACM},
  doi = {10.1145/3077136.3080833},
}

@article{rieh2002judgment,
  title = {Judgment of Information Quality and Cognitive Authority in the Web},
  author = {Rieh, Soo Young},
  journal = {Journal of the American Society for Information Science and Technology},
  volume = {53},
  number = {2},
  pages = {145--161},
  year = {2002},
  doi = {10.1002/asi.10017},
}

@article{rieh2007credibility,
  title = {Credibility: A Multidisciplinary Framework},
  author = {Rieh, Soo Young and Danielson, David R.},
  journal = {Annual Review of Information Science and Technology},
  volume = {41},
  number = {1},
  pages = {307--364},
  year = {2007},
  doi = {10.1002/aris.2007.1440410114},
}

@inproceedings{schwarz2011augmenting,
  title = {Augmenting Web Pages and Search Results to Support Credibility Assessment},
  author = {Schwarz, Julia and Morris, Meredith Ringel},
  booktitle = {Proceedings of the SIGCHI Conference on Human Factors in Computing Systems},
  pages = {1245--1254},
  year = {2011},
  publisher = {ACM},
  doi = {10.1145/1978942.1979127},
}

@misc{gao2024ragsurvey,
  title = {Retrieval-Augmented Generation for Large Language Models: A Survey},
  author = {Gao, Yunfan and Xiong, Yun and Gao, Xinyu and Jia, Kangxiang and Pan, Jinliu and Bi, Yuxi and Dai, Yi and Sun, Jiawei and Wang, Meng and Wang, Haofen},
  year = {2024},
  url = {https://arxiv.org/abs/2312.10997},
  eprint = {2312.10997},
  archiveprefix = {arXiv},
  primaryclass = {cs.CL},
}

@article{ni2026trustworthyrag,
  title = {Towards Trustworthy Retrieval Augmented Generation for Large Language Models: A Survey},
  author = {Ni, Bo and Liu, Zheyuan and Lei, Yongjia and Wang, Leyao and Zhao, Yuying and Cheng, Xueqi and Zeng, Qingkai and Dong, Xin and Xia, Yinglong and Kenthapadi, Krishnaram and Rossi, Ryan and Dernoncourt, Franck and Tanjim, Mehrab and Ahmed, Nesreen and Liu, Xiaorui and Fan, Wenqi and Blasch, Erik and Wang, Yu and Jiang, Meng and Derr, Tyler},
  journal = {ACM Computing Surveys},
  year = {2026},
  doi = {10.1145/3837074},
  url = {https://doi.org/10.1145/3837074},
}

@inproceedings{izacard2021leveraging,
  title = {Leveraging Passage Retrieval with Generative Models for Open Domain Question Answering},
  author = {Izacard, Gautier and Grave, Edouard},
  booktitle = {Proceedings of the 16th Conference of the European Chapter of the Association for Computational Linguistics: Main Volume},
  pages = {874--880},
  year = {2021},
  publisher = {Association for Computational Linguistics},
  doi = {10.18653/v1/2021.eacl-main.74},
}

@inproceedings{yu2024rankrag,
  title = {RankRAG: Unifying Context Ranking with Retrieval-Augmented Generation in LLMs},
  author = {Yu, Yue and Ping, Wei and Liu, Zihan and Wang, Boxin and You, Jiaxuan and Zhang, Chao and Shoeybi, Mohammad and Catanzaro, Bryan},
  booktitle = {Advances in Neural Information Processing Systems},
  volume = {37},
  year = {2024},
  doi = {10.52202/079017-3850},
}

@inproceedings{zhang2025mrag,
  title = {MRAG: A Modular Retrieval Framework for Time-Sensitive Question Answering},
  author = {Zhang, Siyue and Xue, Yuxiang and Zhang, Yiming and Wu, Xiaobao and Luu, Anh Tuan and Zhao, Chen},
  booktitle = {Findings of the Association for Computational Linguistics: EMNLP 2025},
  pages = {3080--3118},
  year = {2025},
  publisher = {Association for Computational Linguistics},
  doi = {10.18653/v1/2025.findings-emnlp.167},
}

@inproceedings{qian2024cfic,
  title = {Grounding Language Model with Chunking-Free In-Context Retrieval},
  author = {Qian, Hongjin and Liu, Zheng and Mao, Kelong and Zhou, Yujia and Dou, Zhicheng},
  booktitle = {Proceedings of the 62nd Annual Meeting of the Association for Computational Linguistics (Volume 1: Long Papers)},
  pages = {1298--1311},
  year = {2024},
  publisher = {Association for Computational Linguistics},
  doi = {10.18653/v1/2024.acl-long.71},
}

@inproceedings{yoon2024compact,
  title = {CompAct: Compressing Retrieved Documents Actively for Question Answering},
  author = {Yoon, Chanwoong and Lee, Taewhoo and Hwang, Hyeon and Jeong, Minbyul and Kang, Jaewoo},
  booktitle = {Proceedings of the 2024 Conference on Empirical Methods in Natural Language Processing},
  pages = {21424--21439},
  year = {2024},
  publisher = {Association for Computational Linguistics},
  doi = {10.18653/v1/2024.emnlp-main.1194},
}

@inproceedings{jin2025longrefiner,
  title = {Hierarchical Document Refinement for Long-context Retrieval-augmented Generation},
  author = {Jin, Jiajie and Li, Xiaoxi and Dong, Guanting and Zhang, Yuyao and Zhu, Yutao and Wu, Yongkang and Li, Zhonghua and Qi, Ye and Dou, Zhicheng},
  booktitle = {Proceedings of the 63rd Annual Meeting of the Association for Computational Linguistics (Volume 1: Long Papers)},
  pages = {3502--3520},
  year = {2025},
  publisher = {Association for Computational Linguistics},
  doi = {10.18653/v1/2025.acl-long.176},
}

@inproceedings{sarthi2024raptor,
  title = {RAPTOR: Recursive Abstractive Processing for Tree-Organized Retrieval},
  author = {Sarthi, Parth and Abdullah, Salman and Tuli, Aditi and Khanna, Shubh and Goldie, Anna and Manning, Christopher D.},
  booktitle = {The Twelfth International Conference on Learning Representations},
  year = {2024},
  url = {https://openreview.net/forum?id=GN921JHCRw},
}

@inproceedings{jiang2023flare,
  title = {Active Retrieval Augmented Generation},
  author = {Jiang, Zhengbao and Xu, Frank and Gao, Luyu and Sun, Zhiqing and Liu, Qian and Dwivedi-Yu, Jane and Yang, Yiming and Callan, Jamie and Neubig, Graham},
  booktitle = {Proceedings of the 2023 Conference on Empirical Methods in Natural Language Processing},
  pages = {7969--7992},
  year = {2023},
  publisher = {Association for Computational Linguistics},
  doi = {10.18653/v1/2023.emnlp-main.495},
}

@misc{yan2024crag,
  title = {Corrective Retrieval Augmented Generation},
  author = {Yan, Shi-Qi and Gu, Jia-Chen and Zhu, Yun and Ling, Zhen-Hua},
  year = {2024},
  url = {https://arxiv.org/abs/2401.15884},
  eprint = {2401.15884},
  archiveprefix = {arXiv},
  primaryclass = {cs.CL},
}

@inproceedings{es2024ragas,
  title = {RAGAs: Automated Evaluation of Retrieval Augmented Generation},
  author = {Es, Shahul and James, Jithin and Espinosa Anke, Luis and Schockaert, Steven},
  booktitle = {Proceedings of the 18th Conference of the European Chapter of the Association for Computational Linguistics: System Demonstrations},
  pages = {150--158},
  year = {2024},
  publisher = {Association for Computational Linguistics},
  doi = {10.18653/v1/2024.eacl-demo.16},
}

@inproceedings{saadfalcon2024ares,
  title = {ARES: An Automated Evaluation Framework for Retrieval-Augmented Generation Systems},
  author = {Saad-Falcon, Jon and Khattab, Omar and Potts, Christopher and Zaharia, Matei},
  booktitle = {Proceedings of the 2024 Conference of the North American Chapter of the Association for Computational Linguistics: Human Language Technologies (Volume 1: Long Papers)},
  pages = {338--354},
  year = {2024},
  publisher = {Association for Computational Linguistics},
  doi = {10.18653/v1/2024.naacl-long.20},
}

@inproceedings{ru2024ragchecker,
  title = {RAGChecker: A Fine-Grained Framework for Diagnosing Retrieval-Augmented Generation},
  author = {Ru, Dongyu and Qiu, Lin and Hu, Xiangkun and Zhang, Tianhang and Shi, Peng and Chang, Shuaichen and Jiayang, Cheng and Wang, Cunxiang and Sun, Shichao and Li, Huanyu and Zhang, Zizhao and Wang, Binjie and Jiang, Jiarong and He, Tong and Wang, Zhiguo and Liu, Pengfei and Zhang, Yue and Zhang, Zheng},
  booktitle = {Advances in Neural Information Processing Systems},
  volume = {37},
  year = {2024},
  doi = {10.52202/079017-0692},
}

@inproceedings{liu2023geval,
  title = {G-Eval: NLG Evaluation using GPT-4 with Better Human Alignment},
  author = {Liu, Yang and Iter, Dan and Xu, Yichong and Wang, Shuohang and Xu, Ruochen and Zhu, Chenguang},
  booktitle = {Proceedings of the 2023 Conference on Empirical Methods in Natural Language Processing},
  pages = {2511--2522},
  year = {2023},
  publisher = {Association for Computational Linguistics},
  doi = {10.18653/v1/2023.emnlp-main.153},
}

@inproceedings{zheng2023judging,
  title = {Judging LLM-as-a-Judge with MT-Bench and Chatbot Arena},
  author = {Zheng, Lianmin and Chiang, Wei-Lin and Sheng, Ying and Zhuang, Siyuan and Wu, Zhanghao and Zhuang, Yonghao and Lin, Zi and Li, Zhuohan and Li, Dacheng and Xing, Eric P. and Zhang, Hao and Gonzalez, Joseph E. and Stoica, Ion},
  booktitle = {Advances in Neural Information Processing Systems},
  volume = {36},
  year = {2023},
  doi = {10.52202/075280-2020},
}

@inproceedings{mu2023gist,
author = {Mu, Jesse and Li, Xiang Lisa and Goodman, Noah},
title = {Learning to compress prompts with gist tokens},
year = {2023},
publisher = {Curran Associates Inc.},
address = {Red Hook, NY, USA},
booktitle = {Proceedings of the 37th International Conference on Neural Information Processing Systems},
articleno = {848},
numpages = {26},
location = {New Orleans, LA, USA},
series = {NIPS '23}
}

\end{document}